\documentclass[preprint]{aastex631}
\usepackage[utf8]{inputenc}
\usepackage{savesym}
\savesymbol{tablenum}
\usepackage{siunitx}
\restoresymbol{SIX}{tablenum}
\usepackage{amsmath}
\usepackage{natbib}
\usepackage{float}
\usepackage{appendix}
\usepackage{multirow}
\usepackage{booktabs}
\usepackage{hyperref}
\usepackage{comment}
\newcommand{\CellWithForceBreak}[2][c]{\begin{tabular}[#1]{@{}c@{}}#2\end{tabular}}

\begin{document}

\title{A High-Resolution Spectroscopic Survey of Directly Imaged Companion Hosts: \\
II. Diversity in C/O Ratios among Host Stars}

\author[0000-0003-3708-241X]{Aneesh Baburaj} \affiliation{Department of Astronomy \& Astrophysics, University of California, San Diego, La Jolla, CA 92093, USA} \affiliation{Department of Physics, University of California, San Diego, La Jolla, CA 92093, USA}

\author[0000-0002-9936-6285]{Quinn M. Konopacky} \affiliation{Department of Astronomy \& Astrophysics, University of California, San Diego, La Jolla, CA 92093, USA} 

\author[0000-0002-9807-5435]{Christopher A. Theissen} \affiliation{Department of Astronomy \& Astrophysics, University of California, San Diego, La Jolla, CA 92093, USA} 

\author[0000-0003-0398-639X]{Roman Gerasimov} \affiliation{Department of Physics and Astronomy, University of Notre Dame, Nieuwland Science Hall, Notre Dame, IN 46556, USA}

\author[0000-0002-9803-8255]{Kielan K. W. Hoch} \affiliation{Space Telescope Science Institute, 3700 San Martin Drive, Baltimore, MD 21218, USA}

\correspondingauthor{Aneesh Baburaj}
\email{ababuraj@ucsd.edu}

\begin{abstract}
 The era of JWST has enabled measurements of abundances of elements such as C, O, and even Na, S, K, and Fe in planetary atmospheres to very high precisions ($\sim$0.1 dex). Accurate inference of planet formation using these elemental abundances require the corresponding abundance measurements for the host star. We present the second set of results from our high-resolution spectroscopic survey of directly imaged companion host stars, measuring abundances of 16 elements (including C, O, Na, Mg, Si, S, K and Fe) for five directly imaged companion host stars. Using both the spectral fitting and the equivalent width methods, we find solar C/O ratios for HR 2562 (0.58 $\pm$ 0.09), AB Pic (0.50 $\pm$ 0.14), and YSES 1 (0.45 $\pm$ 0.05), and sub-solar C/O ratios for PZ Tel (0.28 $\pm$ 0.05) and $\beta$ Pic (0.22 $\pm$ 0.06). The $4\sigma$ sub-solar C/O detections for PZ Tel and $\beta$ Pic highlight the importance of accurate stellar C/O estimates for constraining planet formation. Subsequently, we combine our abundances with those from our previous work to measure population-level average elemental abundances. We find super-solar carbon and oxygen for this stellar population, indicating that the protoplanetary disks around these stars were potentially rich in volatiles. We compare stellar C/O to those of their companions, revealing super-stellar C/O for several objects that suggest planet-like formation mechanisms. We also compare the C/O of our directly imaged companion host star population with other planet host stars using the Kolmogorov-Smirnov Test, which indicates \textbf{insufficient evidence to differentiate between the various stellar populations}.  

\end{abstract}


\section{Introduction}
Direct imaging has led to the discovery of exoplanets characterized by their Jupiter-like masses and separations ranging from 3--5000 au (e.g., \citealt{doi:10.1051/0004-6361/202244727, 2021ApJ...916L..11Z}). The formation of this unique population is not well explained by either of the dominant theories of planet formation -- core/pebble accretion or gravitational instability. The core accretion paradigm struggles as core formation timescales beyond $\sim$ 5 au are too long to reach the critical mass required for runaway gas accretion ($\sim$ 10 $M_{\oplus}$) before the protoplanetary disk dissipates \citep{2009ApJ...707...79D}. While the gravitational instability mechanism can explain planet formation at larger separations, it struggles to form planets $<$ 40 au \citep{2009ApJ...707...79D}. Additionally, the existence of an efficient gravitational instability mechanism would lead to a larger population of brown dwarfs or low-mass stellar companions \citep{2016ARA&A..54..271K}. Large direct imaging surveys have not found evidence for any such population \citep{galicher2016, 2019AJ....158...13N,vigan2021}.

\par Elemental abundance ratios like the atmospheric C/O ratio of Jovian planets has been suggested as a tracer of formation mechanism and location for exoplanets \citep{madhu2011, oberg2011}. A stellar C/O ratio would indicate formation by gravitational instability, or formation inside of both the water snowline and the carbon-grain evaporation line. Meanwhile, super-stellar C/O ratios would indicate that the formation of the atmosphere was dominated by gas accretion beyond the water snowline. This has led to a host of C/O ratio measurements for the direct imaging planets (e.g., \citealt{2013Sci...339.1398K, 2020AJ....160..207W, 2021A&A...648A..59P, palma2023, 2023AJ....166...85H, 2024AA...686A.294C,  2025AJ....169...30B}), as their younger age would preclude significant post-formation atmospheric evolution which can significantly change molecular abundances (e.g., \citealt{zahnle2014}).

\par The degeneracies involved in using the C/O ratio as a formation tracer (e.g., \citealt{2016ApJ...832...41M, 2019ARA&A..57..617M}) has spurred the consideration of other abundance ratios as additional formation tracers. Using pebble-based \citep{sb2021b} and planetesimal-based \citep{2022ApJ...937...36P} core accretion models, \cite{2023ApJ...952L..18C} proposed that volatile-to-sulfur ratios (C/S and O/S) could be used to recover the accretion history of exoplanets . Stellar C/S and O/S ratios would imply formation by planetesimal-based core accretion while super-stellar volatile-to-sulfur ratios would indicate formation by pebble accretion. Highly refractory elements like Na, K, and Fe have also been proposed as formation tracers through the volatile-to-refractory ratio, which in conjunction with the various elemental ratios (C/O, C/S, O/S), can place strong constraints on planet formation \citep{sb2021a,sb2021b}.

\par However, interpretation of these formation tracers in the context of planet formation requires the corresponding abundance ratios for their host stars. Most direct imaging surveys observe young ($<$ 100 Myr) and often, high-mass stars 
($M_*$ $>$ 1.2 $M_{\odot}$) which are are often poorly characterized due to their high rotational velocities ($v\sin{i}$ $>$ 30 kms$^{-1}$) which lead to extensive blending of spectral lines. Even when previous abundance measurements are available, they often lack uncertainty estimates (e.g., AB Pic in \cite{2018AJ....155..111L}) or do not have the entire suite of abundances (C, O, Na, Mg, Si, S, K, Fe, Ni) that may be important in current planet formation studies (e.g., \citealt{2006ApJ...643..484R, oberg2011, 2015A&A...580A..30T}). Several current and impending ground-based and JWST observations seek to measure abundance ratios of exoplanet atmospheres, making it more imperative than ever to have measurements of elemental abundances for the host star (e.g., \citealt{2025AJ....169...30B,hoch2025,2025arXiv251008327M}).

\par In our previous work \cite{2025AJ....169...55B}, henceforth known as Paper I, we presented the first results of an extensive survey of directly imaged companion host stars. Utilizing a combination of spectral template fitting and equivalent widths, we measured carbon and oxygen abundances for five F/G-type directly imaged planet hosts: 51 Eridani, HR 8799, HD 984, GJ 504, and HD 206893. Additionally, we measured the abundances of 13 other elements (Na, Mg, Si, S, Ca, Sc, Ti, Cr, Mn, Fe, Ni, Zn, Y) using the equivalent width method. Our analysis procedure enabled us to obtain elemental abundances to a precision of 0.1--0.2 dex for host stars with $v\sin{i}$ up to 70 kms$^{-1}$ (51 Eridani). For carbon and oxygen, we achieve precisions $\sim$0.1 dex which meets the precision requirement ($\sim$0.1 dex) for constraining exoplanet formation pathways \citep{oberg2011, piso2016, sb2021b}.

\par In this work, we apply the analysis procedure to host stars from our southern hemisphere sample, extending our techniques to a wider range of spectral types as well as higher rotational velocities. The targets chosen include $\beta$ Pic and YSES 1, both primaries of multi-planetary systems that are of great interest to the exoplanet community. Companions to the other targets, HR 2562, AB Pic, and PZ Tel, have all been the subject of extensive study in recent years (e.g., \citealt{palma2023, 2023AJ....165..246F, godoy2024}). We determine their atmospheric properties, metallicities, and abundances of 16 elements (C, O, Na, Mg, Si, S, K, Ca, Sc, Ti, Cr, Mn, Fe, Ni, Zn, Y) up to precisions 0.1--0.2 dex. In Section \ref{sec:targets}, we provide a detailed description of the targets and their companions. Section \ref{sec:obs} presents about our observations and an overview of the data reduction procedure. This is followed by a brief discussion of the methods in Section \ref{sec:methods}, where we focus on the updates to our analysis procedure compared to Paper I. 
The results of our methods are presented in Section \ref{sec:results}. This is followed by the discussion (Section \ref{sec:discuss}) where we compare our results with existing literature and perform a small population analysis using the targets from Paper I and this work. Finally, our results and discussion are summarized in the conclusion (Section \ref{sec:conclusion}).

\section{Science Targets \label{sec:targets}}

\subsection{HR 2562}
HR 2562 is an F5V star hosting a substellar companion, HR 2562 B, at a separation of 20.3 $\pm$ 0.3 au discovered as part of the GPI Exoplanet Survey at Gemini South  \citep{2016ApJ...829L...4K}. The companion has an estimated mass of 30 $\pm$ 15 $M_{Jup}$. The primary star has a range of age estimates in the literature (e.g., \citealt{casa2011, 2013A&A...551L...8P, 2014ApJ...783..121G}) ranging from 0.3--1.6 Gyr. \cite{2016ApJ...829L...4K} assumed an age range of 300--900 Myr to calculate the mass using evolutionary models. \cite{maire2018} subsequently used a combination of astrometric and dynamical analysis to obtain a semi-major axis of 30$^{+11}_{-8}$ au. Recent work includes a dynamical mass upper limit of $\leq$18.5 $M_{Jup}$ by \cite{2023AJ....165..219Z} and an evolutionary model mass of 14 $\pm$ 2 $M_{Jup}$ using the ATMO models \citep{godoy2024}. HR 2562 B has no C/O estimates in the literature.

\subsection{AB Pic}

AB Pic is a young K1V star in the large Tucana-Horologium association, with an estimated age of $\sim$ 30 Myr \citep{2003ApJ...599..342S}. \cite{chauvin2005} discovered AB Pic b, a 13.5 $\pm$ 0.5 $M_{Jup}$ companion (evolutionary model mass) around the primary at a separation of $\sim$200 au \citep{chauvin2005,palma2023}. Since then, AB Pic b has been studied extensively \citep{bonnefoy2010, bonnefoy2014, palma2023, 2025MNRAS.537..134G} with newer evolutionary model mass estimates of 10--14 $M_{Jup}$ by \cite{bonnefoy2010} and C/O ratio of $\sim$ 0.58 from \cite{palma2023} and \cite{2025MNRAS.537..134G}.

\subsection{PZ Tel}

PZ Tel is a young G9IV star in the $\beta$ Pic moving group \citep{2001ApJ...562L..87Z, 2017AJ....154...69S}. A substellar companion (PZ Tel B) was discovered around the primary independently by \cite{2010ApJ...720L..82B}) and \cite{2010A&A...523L...1M}. The former used the Near-Infrared Coronagraphic Imager (NICI; Gemini South) while the latter used VLT/NaCo for their discovery. PZ Tel B has evolutionary model masses ranging from 8--60 $M_{Jup}$ in the literature \citep{2010ApJ...720L..82B, 2012MNRAS.424.1714M, 2014A&A...566A..85S, 2016A&A...587A..56M}. \cite{2023AJ....165..246F} also obtained a dynamical mass measurement of 27$^{+25}_{-9}$ $M_{Jup}$ and semi-major axis of 27$^{+14}_{-4}$ au using joint orbit fit of PZ Tel B incorporating astrometry in the literature, radial velocities, and accelerations from the Hipparcos-Gaia Catalog of Accelerations (HGCA; \citealt{brandt2021}). PZ Tel B has no C/O estimates in the literature.

\subsection{\texorpdfstring{$\beta$}{Beta} Pic}

$\beta$ Pic is a young ($\sim$ 12--21 Myr) A6V star belonging to the $\beta$ Pictoris moving group \citep{1999ApJ...520L.123B, 2001ApJ...562L..87Z, 2014MNRAS.438L..11B}.  The star hosts a well-known debris disk (e.g., \citealt{2003ApJ...584L..27W, 2003ApJ...584L..33W, 2012A&A...542A..40L}), and two planets, $\beta$ Pic b \citep{2009A&A...493L..21L} and $\beta$ Pic c \citep{2019NatAs...3.1135L}. $\beta$ Pic b is at a separation of 9--10 au with a dynamical mass of 9--13 $M_{Jup}$ \citep{2016AJ....152...97W, 2018NatAs...2..883S, 2019NatAs...3.1135L, 2022ApJS..262...21F}. The inner planet $\beta$ Pic c lies at a separation of $\sim$ 2.7 au and has a dynamical mass estimate of 9--11 $M_{Jup}$ \citep{2019NatAs...3.1135L, 2022ApJS..262...21F}. $\beta$ Pic b has previous C/O ratio measurements by \cite{gravity2020} and \cite{landman2024}, which both report C/O $\sim$ 0.4. 

\subsection{TYC 8998-760-1/YSES 1}

TYC 8998-760-1, also known as YSES 1, is a 16.7 $\pm$ 1.4 Myr K3IV star belonging to the Lower Centaurus Crux subgroup of the Scorpius–Centaurus association \citep{pecaut2016,yses1a,yses1b}. The Young Suns Exoplanet Survey (YSES; \citealt{yses1a}) discovered two wide-orbit giant planets around the primary using SPHERE/IRDIS on the VLT \citep{yses1b}. The inner companion YSES 1 b is a 14$\pm$3 $M_{Jup}$ object at a separation of 160 au \citep{yses1a} while the outer YSES 1 c is a 6$\pm$1 $M_{Jup}$  planet at a separation of 320 au \citep{yses1b}. The atmospheres of both YSES 1 b and c have been extensively studied in literature \citep{2021Natur.595..370Z, yses2024}. Most recently, \cite{hoch2025} obtained C/O = 0.62--0.75 for YSES 1 b and C/O = 0.60--0.72 for YSES 1 c.

\section{Observations and Data Reduction \label{sec:obs}}

All targets were observed using the Gemini High-resolution Optical SpecTrograph (GHOST) at Gemini South \citep{2024AJ....168..208K, 2024PASP..136c5001M}. The echelle spectrograph has two arms separated by a dichroic at 530 nm. The blue arm corresponds to wavelengths 347--542 nm, while the red arm covers 520--1060 nm. We utilize the high-resolution mode of GHOST to obtain $R$ $\sim$ 76,000 spectra simultaneously over both arms. To mitigate the effects of chance events (e.g., cosmic rays) impacting our observations, we (1) limit the maximum time per exposure to 900s, and (2) obtain multiple equal exposures for targets with total integration times $\geq$ 900s. The total integration time is chosen such that we obtain SNR $\geq$ 200 per pixel for each of our targets. The exposure times for the targets in this work are outlined in Table \ref{tab:obs}.

\begin{deluxetable}{ccccc}[H]
\tablecaption{Observing schedule for targets in this work\label{tab:obs}}
\tablewidth{0pt}
\tablehead{
\colhead{Target} & \CellWithForceBreak{Number of \\ exposures} & \CellWithForceBreak{Integration Time \\ per exposure (s)} & \CellWithForceBreak{Observation date \\ (UT)} & \CellWithForceBreak{Total Int. \\ Time (s)} 
}
\startdata
{HR 2562}  & {1} & {180}  & {November 4, 2024} & {180} \\ \hline
{AB Pic} & {2} & {750} & {December 7, 2024}   & {1500} \\ \hline 
{PZ Tel}  & {2} & {450}  & {October 2, 2024} & {900} \\ \hline
{beta Pic}  & {1} & {50} & {September 29, 2024} & {50} \\ \hline
{TYC 8998-760-1/}  & {5} & {900} & {April 23, 2025} & \multirow{2}{*}{6300}\\
       {YSES 1}    & {2} & {900} & {April 27, 2025} &  \\ 
\enddata
\end{deluxetable}

We use the DRAGONS (Data Reduction for Astronomy from Gemini Observatory North and South) data reduction pipeline \citep{2023RNAAS...7..214L, dragons_sept_2024}, a platform provided by Gemini Observatory for reducing and processing all Gemini astronomical data for the reduction of our GHOST data. The specific pipeline for the GHOST instrument first applies a bad pixel mask and then bias corrects the science, flats, and arc lamp data (for wavelength calibration) for the red and blue channels. This is followed by flat-fielding of the science and arc lamp data, after which the arc lamp spectra are used for wavelength calibration. We do not apply the spectrophotometric standard for flux calibration or correct the spectra for sky emission lines (tellurics). Tellurics are not removed from our science spectra because we use them to obtain precise wavelength solutions while performing spectral fits. Flux calibration is not required as we perform an order-by-order continuum normalization of the reduced and wavelength-calibrated science spectra before using them for abundance estimations.

\par We adopt the line list from Paper I, for the measurement of elemental abundances. Additionally, we also measure the abundance of potassium (K) as it an important species in the determination of refractory-to-volatile ratios \citep{sb2021b}. Hence, our line lists covers the elements C, O, Na, Mg, Si, S, K, Ca, Sc, Ti, Cr, Mn, Fe, Ni, Zn, and Y. We measure the abundances of carbon and oxygen using both spectral template fitting and equivalent widths, enabling an independent check on the values from each method. For the other 14 elements, we measure their abundances using only the equivalent width method.

\section{Methods \label{sec:methods}}

\subsection{Determination of Temperature, Gravity, and Metallicity}

\par Our work in Paper I showed that model systematics are the dominant source of error in our spectral fits. In Paper I, the adopted noise estimate of 2\% of the median flux value for each spectral order was found to be sufficient to account for the model systematics in our fits. While GHOST spectra come with flux errors, we still choose to adopt the noise estimate similar to our previous work (2\% of the median flux value for each spectral order). All targets in this paper have SNR $\geq$ 200 for their spectra. Hence, our adopted noise estimates are comfortably higher than the actual flux errors and provide a more accurate representation of the spectral fitting errors. 

\par Our preliminary MCMC runs utilize extremely wide prior ranges for $T_\mathrm{eff}\in$ [4500K, 9000K], $\log{g}\in$ [2.9, 5.1], and $\mathrm{[M/H]}\in$ [$-$3, 1], except for the K3IV star YSES 1, for which we use $T_\mathrm{eff} \in$ [4000K, 7000K].  We utilize GHOST order 53 with the Balmer series H$\alpha$ line (whose outer wings are sensitive to $T_\mathrm{eff}$; \citealt{swastik2021}) to obtain an initial estimate of the temperature. Spectral regions around the Na \textsc{i} D doublet (5850--5950 \r{A}, order 58), and the wavelength region of (5600--5750 \r{A}, order 61) and (6100--6200 \r{A}, order 56) are used for obtaining the initial estimates for $\mathrm{[M/H]}$ (e.g., \citealt{petigura2017}). For constraining the telluric parameters, GHOST order 55 is used as it has several telluric features. We only weakly constrain $\log{g}$ in these preliminary runs. The prior ranges for all our parameters is given in Table \ref{tab:priors}. 

\par Once we have initial constraints on $T_\mathrm{eff}$ and $\mathrm{[M/H]}$, and strong constraints on all other parameters (except $\log{g}$), we use an approach that utilizes a combination of single-order and multi-order MCMC runs to obtain the final best-fit values for the stellar $T_\mathrm{eff}$, $\log{g}$, and $\mathrm{[M/H]}$. Each of these MCMC runs involves two rounds of fitting; the first round provides an initial best-fit model, which is then used to reject any $>$ 3$\sigma$ outliers in the data compared to the model. Subsequently, we perform a second round of fitting to obtain the final best-fit parameter values for those orders.

\par The Mg \textsc{i}b triplet (5150--5200 \r{A}) is sensitive to $\log{g}$ (e.g., \citealt{petigura2017, swastik2021}). Hence, we perform a single-order fit on GHOST order 67 with unconstrained $\log{g}$ and the initial guesses for $T_\mathrm{eff}$ and $\mathrm{[M/H]}$. The best-fit values thus obtained are used to restrict the $\log{g}$ prior range in a multi-order fit over six GHOST orders (53, 55, 56, 58, 61, and 67) to further refine the $T_\mathrm{eff}$ and $\mathrm{[M/H]}$. The refined $T_\mathrm{eff}$ and $\mathrm{[M/H]}$ are then used to obtain the best-fit $\log{g}$ by single-order MCMC fits to order 67. This iterative approach requires less computational resources compared to the multiple (often 10+) multi-order fits from Paper I. The number of walkers, total steps, and steps discarded as burn-in in the fitting process depend solely on the number of orders we fit for simultaneously and are specified in Table \ref{tab:num_mcmc}. We also use this new approach on the host star 51 Eridani from Paper I to verify our results. 

\textbf{Since the models do not incorporate uncertainties, our uncertainties are likely underestimated. Achieving high precisions for stellar parameters would need a calibration sample to clearly estimate the systematic uncertainties in the models and methods. However, that is beyond the scope of this work}.

\begin{deluxetable}{ccccc}
\tablecaption{MCMC fitting parameters for different number of orders \label{tab:num_mcmc}}
\tablewidth{0pt}
\tablehead{
\CellWithForceBreak{No. of orders \\ in MCMC run} & \CellWithForceBreak{No. of \\ fitted parameters} & \colhead{No. of walkers} & \colhead{No. of steps} & \colhead{Steps discarded} 
}
\startdata
{1,2}  & {11,13} & {100} & {1000}  & {500} \\ \hline
{3} & {15} & {100} & {1250--1500} & {750--1000}   \\ \hline 
{4}  & {17} & {120} & {2000--2500} & {1500--2000}  \\ \hline
{5}  & {19} & {130} & {2500} & {2000}  \\ \hline
{6}  & {21} & {150} & {3000} & {2500} \\
\enddata
\end{deluxetable}

\subsection{Determination of Carbon and Oxygen Abundances}
\subsubsection{Spectral fitting with synthetic grids \label{subsub:specfit}}
The best-fit stellar atmospheric parameters are used to obtain the stellar carbon and oxygen abundance using spectral template fitting with custom grids that have varying carbon and oxygen abundances. We generate synthetic grids using PySME, the Python wrapper for the spectral analysis software Spectroscopy Made Easy (SME; \citealt{1996A&AS..118..595V, 2017A&A...597A..16P}). With the MARCS stellar atmospheric model grid \citep{marcs2008} and line-lists from the Vienna Atomic Line Database (VALD3; \citealt{1995A&AS..112..525P, 2015PhyS...90e4005R}) as inputs, SME can generate synthetic models for specific values of temperature, gravity, and metallicity. As the MARCS grids span $T_\mathrm{eff}\in$ [2500K, 8000K], $\log{g}\in$ [$-$1, 5], and $\mathrm{[M/H]}\in$ [$-$5, 1], we can obtain custom grids for our targets that cover a small range of temperatures and gravities but have fixed metallicity. To generate synthetic spectra corresponding to the specific set of atmospheric parameters, SME interpolates between the nearest points of the MARCS atmosphere and subsequently runs spectral synthesis with the specified input abundances. The range of carbon and oxygen abundances in the custom grid are specific to each stellar target and are outlined in Table \ref{tab:grids}. \textbf{The grids are created assuming solar abundances from \cite{asplund2009}.}

\par The carbon (C \textsc{i}) features in order 48 (7111--19 \r{A}), 53 (6587 \r{A}), 64 (5380 \r{A}), 68 (5052 \r{A}), 70 (4930 \r{A}), and 72 (4772 \r{A}) are used to determine the carbon abundance. However, depending on the strength of the C \textsc{i} feature and the model fits to the different carbon orders, not all orders are used for each target. To determine the oxygen abundance, we utilize the oxygen triplets at 6155--58 \r{A} (order 56) and 7771--75 \r{A} (order 44). For the slow-rotating targets, the weak O \textsc{i} forbidden line at 6300 \r{A} (order 55) can also be used to obtain the oxygen abundance. For both elements, multi-order fits are used to determine the final best-fit abundance. As the oxygen triplet at 7771--75 \r{A} has significant non-equilibrium local thermodynamic (NLTE) effects, we use the NLTE corrections from \cite{amarsi2019}, which are fixed offsets as functions of $T_\mathrm{eff}$ and $\log{g}$. We also apply NLTE corrections to the oxygen triplet at 6155--6158 for HR 2562 and $\beta$ Pic as the corrections are $\sim -0.1$ dex for these targets. \textbf{We validate the VALD line-lists by synthesizing models with varying carbon and oxygen abundances for the paper I host star 51 Eridani and performing spectral fits. We recover the same abundances we obtain in paper I within 1$\sigma$ uncertainties (Table \ref{tab:litcompare})}. 

\par The estimated temperature for $\beta$ Pic does not fall within the parameter space of the MARCS grid. For this specific target, we instead create a grid using the ATLAS9 stellar atmospheric models \citep{2002A&A...392..619H,2017ascl.soft10017K}. Additionally, $\beta$ Pic has an unusual abundance profile \textbf{with a solar metallicity ([M/H] = 0) from the atmospheric parameter spectral fitting procedure, but sub-solar abundances of several elements, including iron. As the orders used to measure metallicity have spectral lines due to various elements, we do not get [M/H] $<$ 0 if [X/H] $<$ 0 for only a few of the elements}. Hence, while creating the synthetic grid we set the abundances of Mg, Si, Sc, Cr, Mn, Fe, and Ni to the values determined using equivalent widths (Table \ref{tab:ewabundance}). The rest of our analysis for both the MARCS and ATLAS synthetic grids (henceforth called \textit{MARCS-C/O} and \textit{ATLAS-C/O} grids) is same as that for the \textit{PHOENIX-C/O} grids from Paper I.

\subsection{Determination of Equivalent Widths}
The equivalent width approach is used to obtain abundances of carbon, oxygen, and 14 other elements (Na, Mg, Si, S, K, Ca, Sc, Ti, Cr, Mn, Fe, Ni, Zn, Y). We use the spectral analysis software MOOG \citep{1973ApJ...184..839S} and the Kurucz-ATLAS9 models generated using the BasicATLAS framework \citep{2023AJ....165....2L} for converting equivalent width measurements to the corresponding abundances for an atomic species. Since our spectra are already continuum normalized during data reduction, we do not need to perform additional continuum normalization. For lines that have severe rotational blending, trapezoidal integration is used to obtain the equivalent width of the entire blended feature. Subsequently, the `blends' driver in MOOG is used to obtain the abundance of the desired species. The `blend' driver uses the input Kurucz-ATLAS9 model to calculate equivalent widths for the other lines in the blend. The residual equivalent width corresponds to that of the line of interest (i.e., the desired species), which is then used to calculate the corresponding abundance. For atomic lines with no/minor blending, where the shape of the particular line is clearly visible, we fit a Gaussian profile to the line followed by trapezoidal integration on the best-fit Gaussian profile to obtain the equivalent widths. The `abfind' driver in MOOG is then used to force-fit abundances to the equivalent widths of these single lines. For all our targets we consider three sources of abundance uncertainties that are combined in quadrature: 1) line-to-line uncertainty, 2) uncertainty in $T_\mathrm{eff}$, and 3) uncertainty in $\log{g}$. Hence, the final abundance uncertainty for each species takes into account line-to-line scatter, as well as the uncertainties in $T_\mathrm{eff}$ and $\log{g}$. As we calculate \textit{absolute} abundances, the [M/H] has a minor effect on the result, allowing us to only consider error contributions from $T_\mathrm{eff}$ and $\log{g}$. \textbf{We define absolute abundance along the lines of \cite{asplund2009} and is equivalent to $\log{\epsilon_X}$, which is:
\begin{equation}
    \log{\epsilon_X} = \log{X} - \log{H} + 12
\end{equation}}. 
Additionally, for oxygen, we also take into account uncertainties in 3D/1D NLTE to 1D LTE corrections, \textbf{which are obtained from \cite{amarsi2019} \footnote{The specific IDL tools were provided by the authors and can be obtained from \href{https://www.astro.uu.se/~amarsi/cofe_tools.tar.gz}{https://www.astro.uu.se/$\sim$amarsi/cofe$\_$tools.tar.gz}}}.

\par As $\beta$ Pic is an extremely fast rotator ($v\sin{i}$ $\sim$ 115 kms$^{-1}$) and has unusual abundances only for specific elements, using the blends driver directly does not provide accurate abundances. Instead, for each ionic species, we first use the abfind driver on spectral lines which have negligible blend contributions from spectral lines of other species to obtain an initial abundance estimate for that species. This abundance estimate is then used to resolve a blended feature which has significant contributions (EW $>$ 0.5 \r{A}) from spectral lines of different species. In this way, we calculate the equivalent width of our line of interest in the blended feature, which is then converted to the corresponding abundance using MOOG.

\section{Results \label{sec:results}}
\subsection{HR 2562}
\begin{figure}
    \centering
    \includegraphics[width=1.0\linewidth]{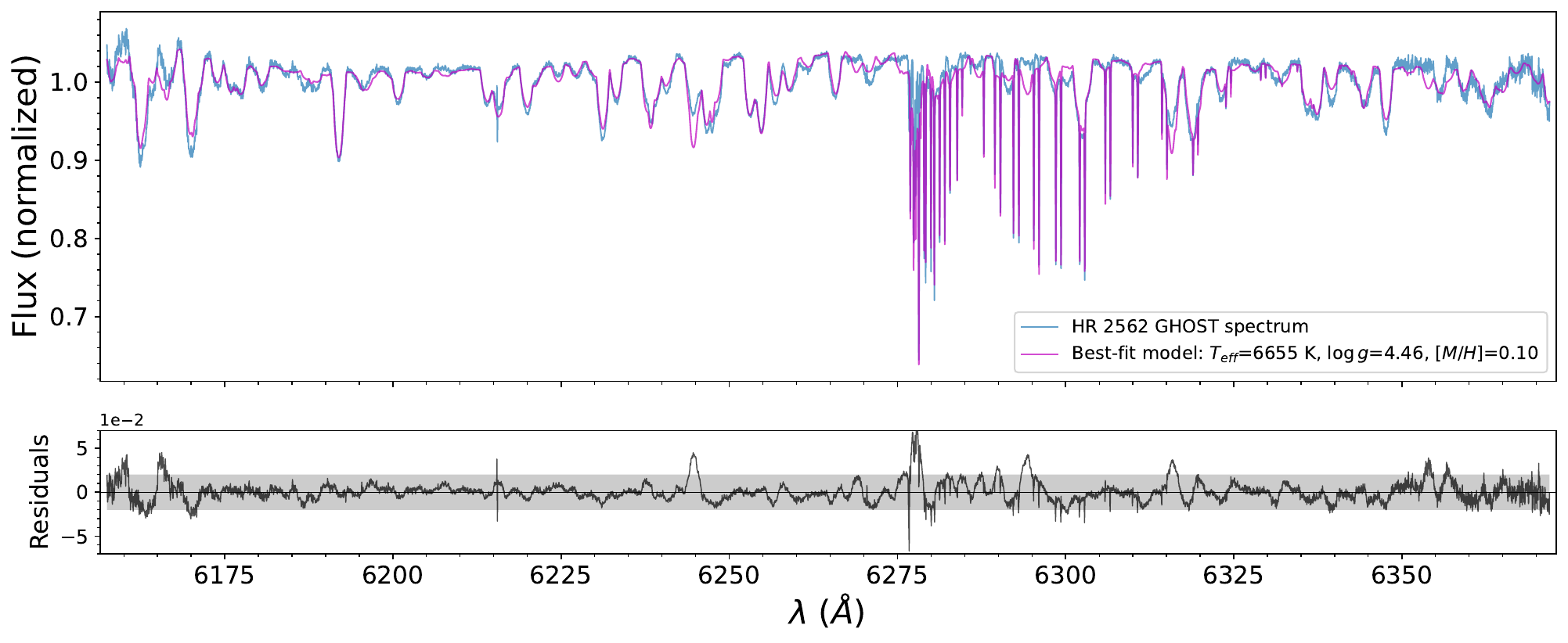}
    \caption{Best-fit PHOENIX model to the APF spectrum for the target HR 2562 (cyan), shown for GHOST order 55. The broader absorption features correspond to stellar features while the sharper ones (those similar to spikes) correspond to telluric features. This model has $T_\mathrm{eff}$ = 6655 K, $\log{g}$ = 4.46, $\mathrm{[M/H]}$ = 0.10 (magenta). The residuals between the data and the model are plotted in black and other noise limits are shown in gray. }
    \label{fig:hr2562spec55}
\end{figure}

\begin{figure}
    \centering
    \includegraphics[width=0.8\linewidth]{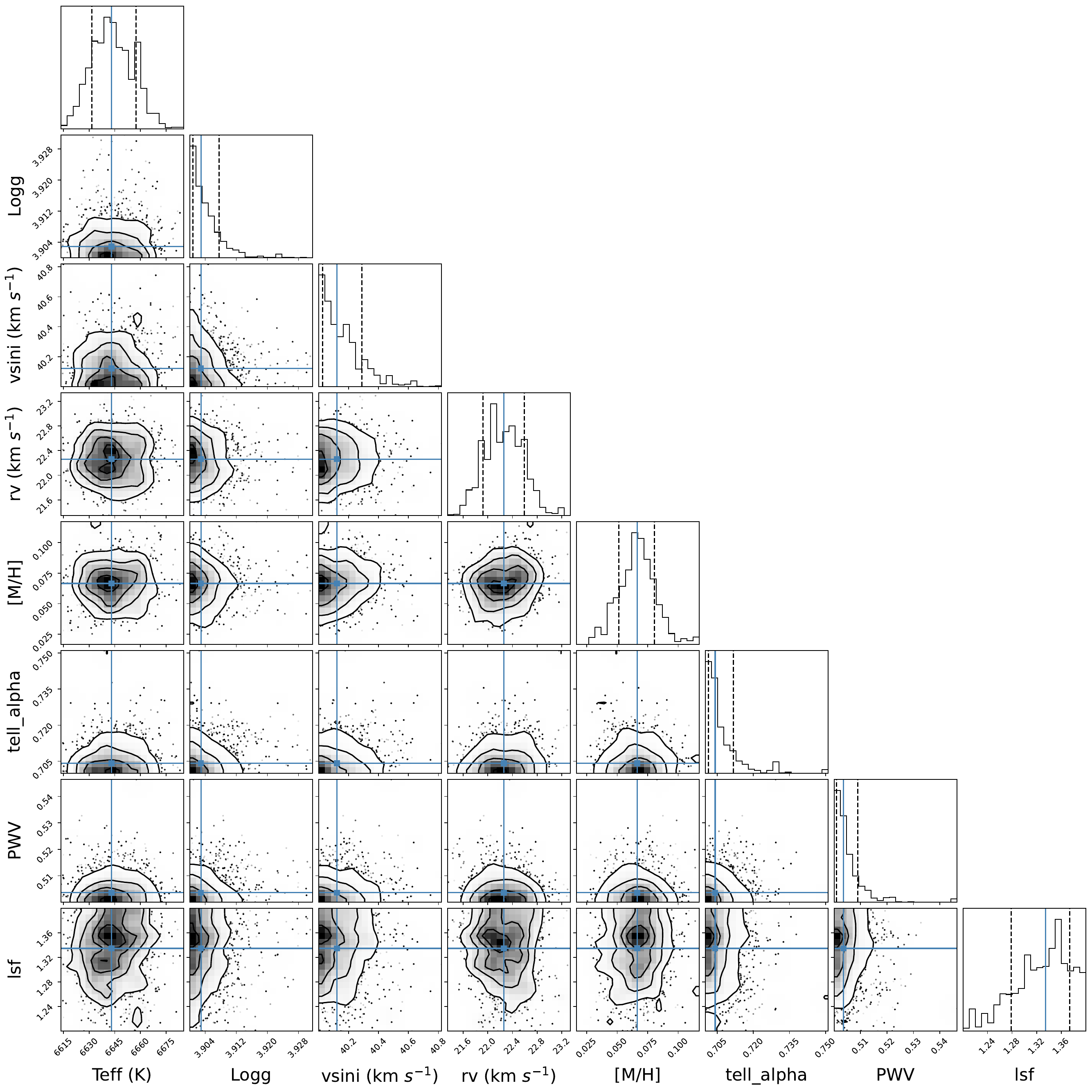}
    \caption{Corner plot for a single-order MCMC run fitting order 53 (with the H$\alpha$ feature) of the spectrum of HR 2562. Although some parameters are not well constrained (e.g., $\log{g}$), we use this order only to determine the best-fit temperature. The marginalized posteriors are shown along the diagonal. The blue lines represent the 50 percentile, and the dotted lines represent the 16 and 84 percentiles. The subsequent covariances between all the parameters are in the corresponding 2-D histograms. This MCMC run gives best-fit $T_\mathrm{eff}$ = 6644 $\pm$ 14 K.}
    \label{fig:cornerhr2562atmos}
\end{figure}

The combination of single-order and multi-order fits for HR 2562 gave $T_\mathrm{eff}$ = 6655 $\pm$ 45 K, $\log{g}$ = 4.46 $\pm$ 0.19, $\mathrm{[M/H]}$ = 0.10 $\pm$ 0.05 as the stellar atmospheric parameters. Figure \ref{fig:hr2562spec55} and \ref{fig:cornerhr2562atmos} show the best-fit model to spectral fit data and the posteriors from the a single-order MCMC run used to determine best-fit temperature, respectively. All spectra shown in this work are in air and in the observed frame of reference.

\par
\begin{figure}
    \centering
    \plottwo{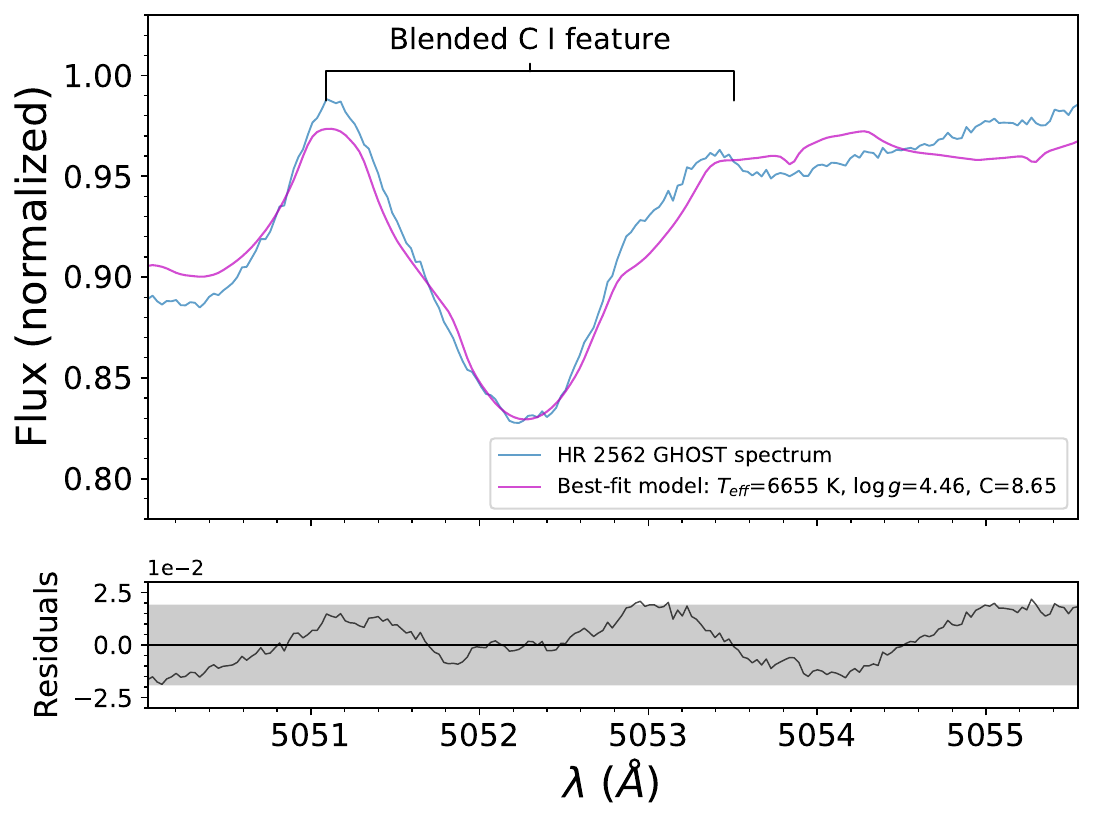}{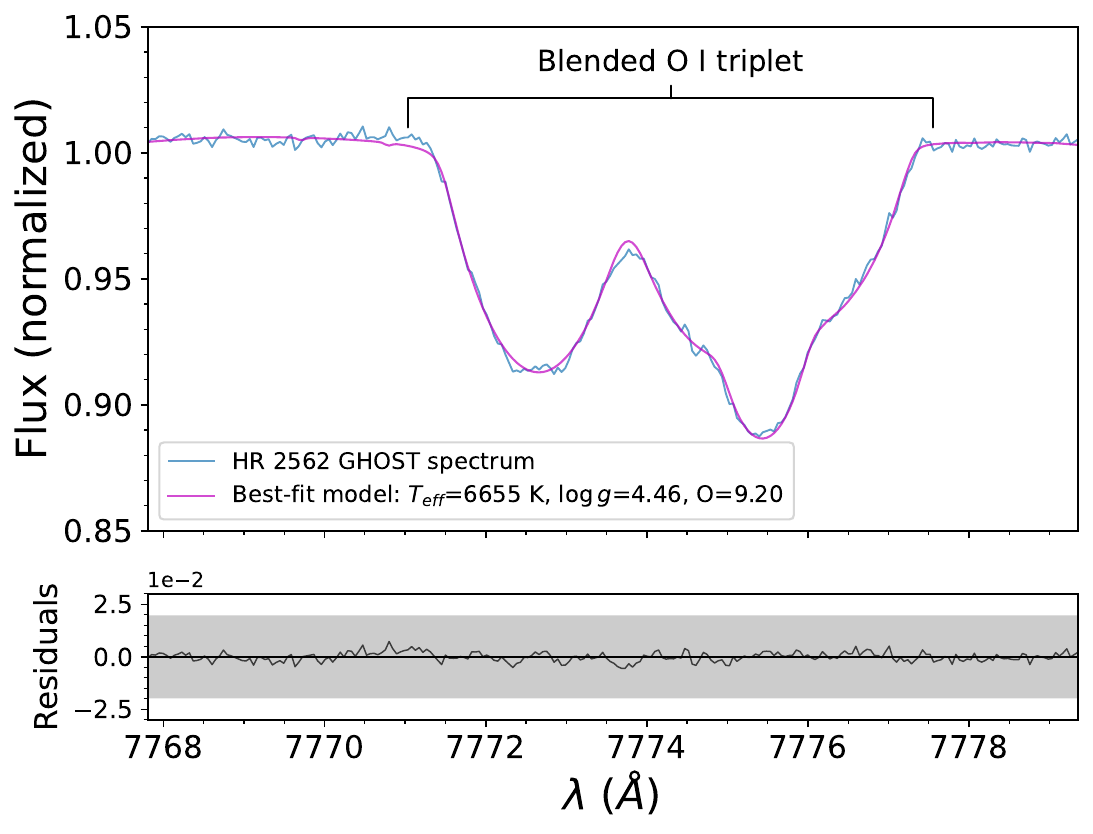}
    \caption{Best-fit \textit{MARCS-C/O} model (magenta) to the GHOST spectrum of the star HR 2562 (cyan), shown for (\textit{Left}) the blended C \textsc{i} line at 5052 \r{A} and (\textit{Right}) the blended O \textsc{i} triplet feature at 7771--75 \r{A}. The residuals between the data and the model are plotted in black and other noise limits are shown in grey. Spectral fit of the \textit{MARCS-C/O} grid to the C \textsc{i} features give best-fit $\log{\epsilon_C}$ = 8.65. Spectral fit to the 7771--75 \r{A} O \textsc{i} triplet gives a best-fit $\log{\epsilon_O}$ = 9.20. After applying NLTE corrections, we get $\log{\epsilon_O}$ = 8.86.}
    \label{fig:hr2562specco}
\end{figure}

\begin{figure}
    \centering
    \includegraphics[width=0.8\linewidth]{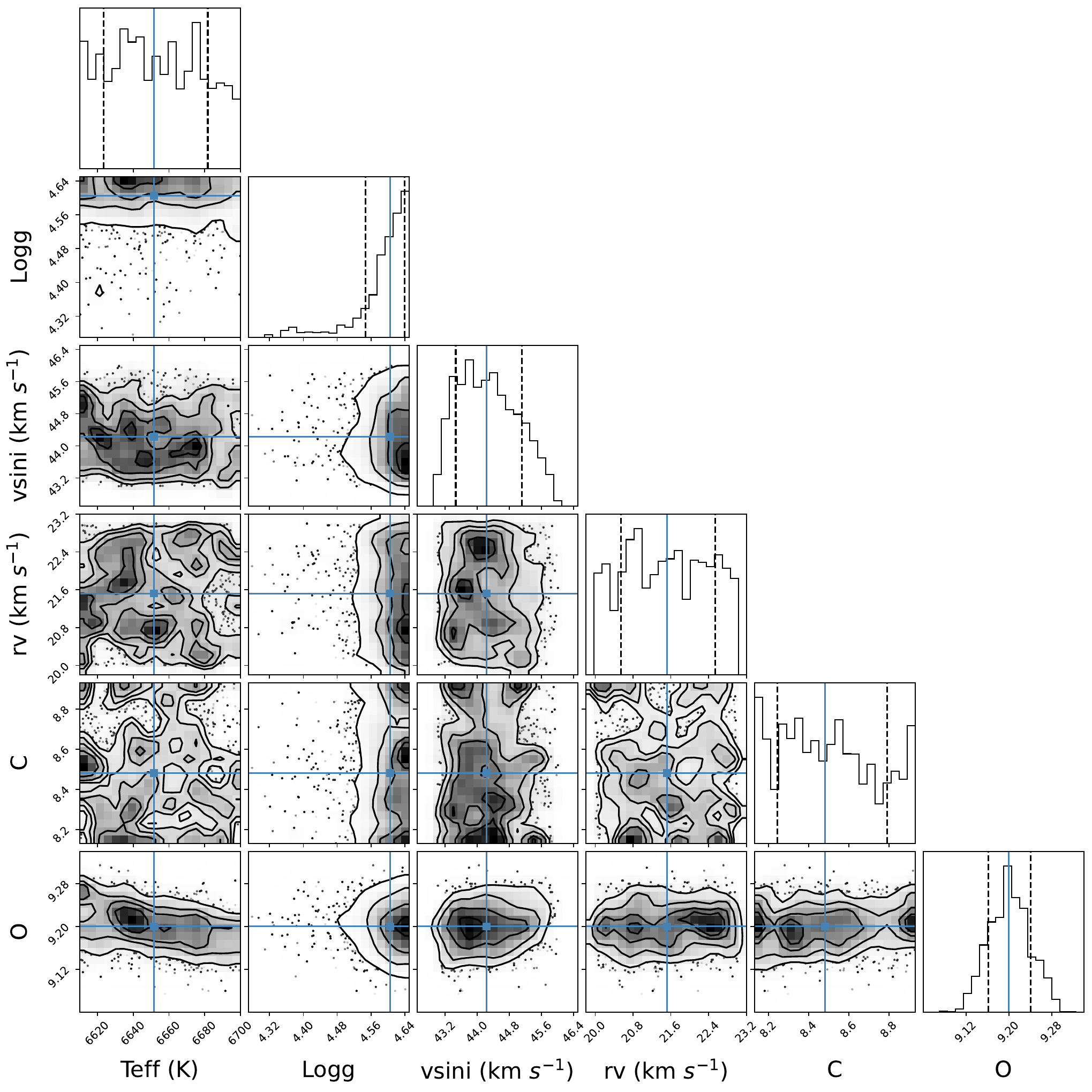}
    \caption{Corner plot for \textit{MARCS}--\textit{C/O} grid fit to spectral order with the O \textsc{i} triplet at 7771--75 \r{A}. The marginalized posteriors are shown along the diagonal. The blue lines represent the 50 percentile, and the dotted lines represent the 16 and 84 percentiles. The subsequent covariances between all the parameters are in the corresponding 2-D histograms. We obtain a best-fit $\log{\epsilon_O}$ = 9.20 $\pm$ 0.04, corresponding to $\log{\epsilon_O}$ = 8.86 $\pm$ 0.04 after NLTE correction. The carbon is not well-constrained fitting this order.}
    \label{fig:cornerhr2562co}
\end{figure}

 The MARCS-C/O grid with the best-fit metallicity of 0.1 was used to determine the carbon and oxygen abundances. We simultaneously fit three carbon orders using 110 walkers and 1500 steps, with the first 1000 steps discarded as burn-in. Our spectral fits give [C/H] = 0.22 $\pm$ 0.05. Due to differing NLTE corrections, we perform independent MCMC fits on the oxygen triplet at 6165--68 \r{A} and 7771--75 \r{A} using MCMC runs of 100 walkers, 1000 steps with the first 500 steps discarded.  After applying NLTE corrections from \cite{amarsi2019}, we get [O/H] = 0.20 $\pm$ 0.05, giving us a spectral fit C/O = 0.58 $\pm$ 0.09.
Figure \ref{fig:hr2562specco} (\textit{Right}) and Figure \ref{fig:cornerhr2562co} show the best-fit MARCS-C/O model to the 7771--75 \r{A} triplet and the posterior plots respectively.

\par Using the equivalent width approach gives us [C/H] = 0.07 $\pm$ 0.14, [O/H] = 0.04 $\pm$ 0.10, and [S/H] = 0.20 $\pm$ 0.13. These lead to C/O = 0.59 $\pm$ 0.23, C/S = 15.14 $\pm$ 6.66, and O/S = 25.70 $\pm$ 9.71. Thus, the C/O from both spectral fitting and equivalent width agree with each other as well as with the solar value (0.55 $\pm$ 0.09; \citealt{asplund2009}).

\subsection{AB Pic}
The stellar atmospheric parameters obtained for AB Pic are $T_\mathrm{eff} = 5006 \pm 81$~K, $\log{g} = 4.40 \pm 0.09$, and $\mathrm{[M/H]} = -0.07 \pm 0.05$. We used a MARCS-C/O grid with the best-fit metallicity = -0.07 to obtain the carbon and oxygen abundances. For carbon, we fit four orders simultaneously with MCMC runs of 130 walkers and 2000 steps, with the first 1500 discarded as burn-in. This gives us [C/H] $= 0.05 \pm 0.10$. The oxygen abundance was determined using the forbidden O \textsc{i} line at 6300 \r{A}, and the O \textsc{i} triplet at 7771--75 \r{A}. Both orders were fit independently (due to possible NLTE corrections) with runs of 100 walkers, 750 steps, and 500 discarded as burn-in, giving a final [O/H] $= 0.09 \pm 0.07$. Thus, we obtain a spectral fit C/O $= 0.50 \pm 0.14$.

\par By measuring the equivalent widths, we obtain [C/H] = 0.05 $\pm$ 0.13, [O/H] = 0.22 $\pm$ 0.20, and [S/H] = 0.26 $\pm$ 0.19, which give C/O = 0.37 $\pm$ 0.20, C/S = 12.59 $\pm$ 6.67, O/S = 33.88 $\pm$ 21.52. The spectral-fit and equivalent width C/O agree with each other and are effectively solar.

\subsection{PZ Tel}
We obtain a $T_\mathrm{eff} = 5014 \pm 63$~K, $\log{g}$ = 4.05 $\pm$ 0.15, and $\mathrm{[M/H]} = -0.17 \pm 0.06$ for PZ Tel from our stellar atmosphere MCMC fits. The MARCS-C/O grid with the best-fit metallicity of $-0.17$ was used to determine abundances of carbon and oxygen. We fit the four orders using 130 walkers, 2500 steps and the first 2000 discarded as burn-in, giving us [C/H] $= -0.04 \pm 0.07$. For oxygen, only the oxygen triplet at 7771--75 \r{A} could be used for abundance measurements. MCMC runs with 100 walkers, 1000 steps and the first 500 discarded as burn-in gave a best-fit [O/H] = $0.26 \pm 0.04$ after NLTE corrections. The error here includes the error in NLTE corrections. Thus, we get a super-solar oxygen abundance at $>5\sigma$, which translates to a sub-solar C/O = $0.28 \pm 0.05$ at $4\sigma$. 

\par Equivalent widths of spectral lines give us [C/H] = $0.02 \pm 0.21$, [O/H] =  $0.42 \pm 0.12$, and [S/H] = $0.19 \pm 0.23$, leading to C/O = 0.22 $\pm$ 0.12, C/S = 13.80 $\pm$ 9.90, and O/S = 63.10 $\pm$ 37.69. The super-solar oxygen and sub-solar C/O from this approach are consistent with the spectral fit C/O. The oxygen deviates from solar at 3.5$\sigma$, while the C/O deviates from solar C/O (0.55 $\pm$ 0.09) at 2$\sigma$. 

\subsection{\texorpdfstring{$\beta$}{Beta} Pic}
The atmospheric MCMC fits give $T_\mathrm{eff} = 8054 \pm 96$~K, $\log{g}$ = 4.18 $\pm$ 0.13, and $\mathrm{[M/H]}$ = 0.01 $\pm$ 0.03.  Carbon and oxygen abundances are determined using an ATLAS-C/O grid with [M/H] = 0.01 and custom abundances of certain elements (refer Section \ref{subsub:specfit}). For carbon, we fit five orders simultaneously using MCMC runs of 2000 steps, 140 walkers with 1500 burn-in steps, giving us a [C/H] $= -0.25 \pm 0.06$. In order to be conservative with our estimates, we round the uncertainty to 0.10 dex, which leads to it being sub-solar at $>2\sigma$. 

 \par \cite{amarsi2019} do not have 3D NLTE to 1D LTE corrections available for stars at such high temperatures. Hence, as the O \textsc{i} triplet at 7771--75 \r{A} has much higher NLTE corrections ($>$0.5 dex for stars with $T_\mathrm{eff}$ $>$ 6400 K and solar metallicities), we use only the O \textsc{i} triplet at 6155--58 \r{A} for our analysis. For the 6155--58 \r{A} triplet, we adopt the 1D NLTE to 1D LTE correction of $-0.09$ dex. We perform MCMC fits with 100 walkers and 1000 steps (of which 500 were discarded). After applying the NLTE correction, we get [O/H] = 0.14 $\pm$ 0.05 for 6155--58 \r{A} triplet, giving spectral fit C/O = 0.22 $\pm$ 0.06. The [O/H] is super-solar at 2$\sigma$ while the C/O is sub-solar at a $4\sigma$ significance.

\par The equivalent width method gives us [C/H] = -0.27 $\pm$ 0.10 and [O/H] = 0.14 $\pm$ 0.07 for the 6155--58 \r{A} triplet. We get a C/O = 0.21 $\pm$ 0.06, which agrees with the spectral fit value and is sub-solar at 4$\sigma$. We get [S/H] = 0.17 $\pm$ 0.14, leading to C/S = 7.41 $\pm$ 2.94, and O/S = 34.67 $\pm$ 12.50. The equivalent width gives super-solar oxygen, and sub-solar carbon and C/O, all of which are in good agreement with the spectral fit values.

\subsection{YSES 1}
YSES 1 shows signatures of H$\alpha$ emission due to the presence of accreting planets \citep{hoch2025}, ruling out the use of order 53 (with the H$\alpha$ feature) for determination of stellar parameters. We instead resort to the order 58 with the Na \textsc{i} D doublet and other spectral features for temperature determination (e.g., \citealt{swastik2021}). The rest of the procedure is similar to other targets. We finally get $T_\mathrm{eff} = 4709 \pm 45$~K, $\log{g}$ = 4.28 $\pm$ 0.09, and $\mathrm{[M/H]} = -0.02 \pm 0.03$. A MARCS-C/O grid with [M/H] $= -0.02$ was used to obtain the carbon and oxygen abundances. For carbon, we simultaneously fit three orders using MCMC runs of 1500 steps, 110 walkers and the first 1000 discarded as burn-in, giving [C/H] $= -0.04 \pm 0.04$. Oxygen involves independently fitting the 6300 \r{A} O \textsc{i} forbidden line and O \textsc{i} triplet at 7771--75 \r{A} due to different NLTE corrections involved. Both sets of MCMC fits involve 100 walkers, 1000 steps with 500 discarded as burn-in, ultimately giving [O/H] $= 0.05 \pm 0.03$. Thus, we get a spectral fit C/O = $0.45 \pm 0.05$.

\par Equivalent widths give us [C/H] = $0.01 \pm 0.14$, [O/H] = $0.14 \pm 0.14$, and [S/H] = $0.36 \pm 0.20$, leading to a C/O = $0.41 \pm 0.19$, C/S = $9.12 \pm 5.13$, and O/S = $22.39 \pm 12.55$. The C/O agrees excellently with the spectral fit value and is consistent with solar.

\begin{deluxetable}{lcccccccccccccccc}
\rotate
\tabletypesize{\footnotesize} 
\tablecaption{Equivalent width elemental abundance corresponding to various spectral species\label{tab:ewabundance}}
\tablewidth{0pt}
\tablehead{
\colhead{Species} & \colhead{HR 2562} & \colhead{} & \colhead{} & \colhead{AB Pic} & \colhead{} & \colhead{} & \colhead{PZ Tel} & \colhead{} & \colhead{} & \colhead{$\beta$ Pic} & \colhead{} & \colhead{} & \colhead{YSES 1} & \colhead{} & \colhead{} & \colhead{Solar values} \\
\colhead{} & 
\colhead{log(N)} & \colhead{$\sigma$} & \colhead{\# lines}  &
\colhead{log(N)} & \colhead{$\sigma$} & \colhead{\# lines}  &
\colhead{log(N)} & \colhead{$\sigma$} & \colhead{\# lines}  &
\colhead{log(N)} & \colhead{$\sigma$} & \colhead{\# lines}  &
\colhead{log(N)} & \colhead{$\sigma$} & \colhead{\# lines} & \colhead{}
}
\startdata
C \textsc{i}   & 8.50 & 0.14 & 5 & 8.48 & 0.13 & 3 & 8.45 & 0.21 & 3 & 8.16 & 0.10 & 3 & 8.44 & 0.14 & 2 & 8.43 \\
O \textsc{i}   & 8.73 & 0.10 & 2\tablenotemark{\footnotesize{a}} & 8.91 & 0.20 & 4 & 9.11 & 0.12 & 1\tablenotemark{\footnotesize{a}} & 8.83 & 0.07 & 2\tablenotemark{\footnotesize{a}} & 8.83 & 0.14 & 2 & 8.69 \\
Na \textsc{i}  & 6.48 & 0.05 & 1 & 6.24 & 0.08 & 1 & 5.93 & 0.06 & 1 & 6.23 & 0.05 & 1 & 6.23 & 0.06 & 1 & 6.24 \\
Mg \textsc{i}  & 7.64 & 0.13 & 3 & 7.63 & 0.11 & 2 & 7.29 & 0.09 & 2 & 7.37 & 0.14 & 4 & 7.65 & 0.15 & 3 & 7.60 \\
Si \textsc{i}  & 7.69 & 0.19 & 4 & 7.59 & 0.11 & 11 & 7.42 & 0.07 & 2 & 7.33 & 0.05 & 1 & 7.54 & 0.12 & 11 & 7.51 \\
Si \textsc{ii} & 7.79 & 0.11 & 2 & 7.98\tablenotemark{\footnotesize{b}} & 0.15 & 2 & 7.45 & 0.13 & 2 & 7.77\tablenotemark{\footnotesize{b}} & 0.20 & 2 & 8.01\tablenotemark{\footnotesize{b}} & 0.11 & 2 & 7.51 \\
S \textsc{i}   & 7.32 & 0.13 & 2 & 7.38 & 0.19 & 2 & 7.31 & 0.23 & 2 & 7.29 & 0.14 & 2 & 7.48 & 0.20 & 2 & 7.12 \\
K \textsc{i}   & 5.22 & 0.03 & 1 & 5.08 & 0.12 & 2 & 4.90 & 0.10 & 1 & 5.11 & 0.07 & 1 & 5.07 & 0.09 & 2 & 5.03 \\
Ca \textsc{i}  & 6.46 & 0.09 & 7 & 6.39 & 0.15 & 9 & 6.04 & 0.13 & 10 & 6.31 & 0.20 & 11 & 6.35 & 0.15 & 14 & 6.34 \\
Sc \textsc{ii} & 3.17 & 0.07 & 2 & 3.18 & 0.09 & 5 & 2.89 & 0.13 & 3 & 2.98 & 0.05 & 1 & 3.16 & 0.08 & 6 & 3.15 \\
Ti \textsc{i}  & 5.25 & 0.10 & 4 & 4.94 & 0.25 & 7 & 4.54 & 0.09 & 1 & 4.95 & 0.10 & 2 & 5.07 & 0.10 & 8 & 4.95 \\
Ti \textsc{ii} & 5.28 & 0.17 & 2 & 5.06 & 0.12 & 6 & 4.70 & 0.06 & 2 & 4.91 & 0.10 & 3 & 5.00 & 0.12 & 9 & 4.95 \\
Cr \textsc{i}  & 5.74 & 0.11 & 1 & 5.62 & 0.17 & 9 & 5.28 & 0.23 & 3 & 5.52 & 0.08 & 2 & 5.62 & 0.09 & 9 & 5.64 \\
Cr \textsc{ii} & 5.77 & 0.16 & 8 & 5.80 & 0.14 & 6 & 5.44 & 0.12 & 4 & 5.47 & 0.05 & 5 & 5.67 & 0.14 & 5 & 5.64 \\
Mn \textsc{i}  & 5.54 & 0.05 & 3 & 5.44 & 0.07 & 4 & 4.97 & 0.19 & 2 & 5.12 & 0.11 & 2 & 5.50 & 0.11 & 5 & 5.43 \\
Fe \textsc{i}  & 7.64 & 0.16 & 55 & 7.47 & 0.17 & 69 & 7.31 & 0.16 & 70 & 7.26 & 0.20 & 42 & 7.47 & 0.11 & 75 & 7.50 \\
Fe \textsc{ii} & 7.66 & 0.13 & 12 & 7.59 & 0.18 & 11 & 7.35 & 0.18 & 14 & 7.29 & 0.19 & 13 & 7.57 & 0.16 & 12 & 7.50 \\
Ni \textsc{i}  & 6.43 & 0.15 & 8 & 6.25 & 0.10 & 9 & 5.98 & 0.16 & 4 & 6.05 & 0.13 & 3 & 6.21 & 0.09 & 9 & 6.22 \\
Zn \textsc{i}  & 4.66 & 0.07 & 3 & 4.57 & 0.04 & 2 & 3.60 & 0.21 & 2 & 4.54 & 0.15 & 2 & 4.56 & 0.08 & 2 & 4.56 \\
Y \textsc{ii}  & 2.41 & 0.08 & 2 & 2.21 & 0.05 & 3 & 2.08 & 0.07 & 2 & 2.24 & 0.09 & 3 & 2.37 & 0.20 & 3 & 2.21 
\enddata
\tablenotetext{}{All abundances are given as absolute [X/H] + 12}
\tablenotetext{a}{O I triplets blended into a single feature are counted as singular lines}
\tablenotetext{b}{Abundance possibly corresponds to equivalent width of unresolved blend}
\end{deluxetable}

\begin{deluxetable}{c|c|ccccc}
\tabletypesize{\footnotesize} 
\tablecaption{Literature comparisons \label{tab:litcompare}}
\tablewidth{0pt}
\tablehead{\colhead{Target}    & \colhead{Work}     & \colhead{$T_\mathrm{eff}$ (K)}    & \colhead{$\log{g}$ (cgs)}     & \colhead{[M/H]}     & \colhead{[C/H]} & \colhead{[O/H]}
}
\startdata
\multirow{6}{*}{HR 2562}  & \multirow{2}{*}{This work} & \multirow{2}{*}{6655 $\pm$ 45} & \multirow{2}{*}{4.46 $\pm$ 0.19} & \multirow{2}{*}{0.10 $\pm$ 0.05} & 0.22 $\pm$ 0.05 \tablenotemark{a} &  0.20 $\pm$ 0.05 \tablenotemark{a }\\ 
  & & & & & 0.07 $\pm$ 0.14 \tablenotemark{b} & 0.04 $\pm$ 0.10 \tablenotemark{b} \\ 
  & \cite{zak2022} & 6603 $\pm$ 44 & 4.34 & 0.24 & & \\ 
  & \cite{swastik2021} & 6785 $\pm$ 29 & 4.40 $\pm$ 0.05 & 0.21 $\pm$ 0.03 & & \\
  & \cite{mesa2018} & 6650 $\pm$ 100 & 4.3 $\pm$ 0.2 & 0.13 $\pm$ 0.02 &  &  \\  
  & \cite{casa2011} & 6597 $\pm$ 81 & 4.28 & 0.08 &  &  \\ \hline
\multirow{4}{*}{AB Pic\tablenotemark{c}}  & \multirow{2}{*}{This work} & \multirow{2}{*}{5006 $\pm$ 81} & \multirow{2}{*}{4.40 $\pm$ 0.09} & \multirow{2}{*}{-0.07 $\pm$ 0.05} & 0.05 $\pm$ 0.10 \tablenotemark{a} &  0.09 $\pm$ 0.07 \tablenotemark{a }\\ 
  & & & & & 0.05 $\pm$ 0.13 \tablenotemark{b} & 0.22 $\pm$ 0.20 \tablenotemark{b} \\ 
  & \cite{2022ApJS..259...35A} & 4750 $\pm$ 13 & 4.50 $\pm$ 0.02 & -0.09 $\pm$ 0.01 & -0.09 $\pm$ 0.02 & -0.06 $\pm$ 0.01  \\
  & \cite{2018AJ....155..111L} & 5030 $\pm$ 80 & 4.47 & 0.09 $\pm$ 0.10 & -0.15 & 0.02 \\ \hline
\multirow{5}{*}{PZ Tel}  & \multirow{2}{*}{This work} & \multirow{2}{*}{5014 $\pm$ 63} & \multirow{2}{*}{4.05 $\pm$ 0.15} & \multirow{2}{*}{-0.17 $\pm$ 0.06} & -0.04 $\pm$ 0.07 \tablenotemark{a} &  0.26 $\pm$ 0.04 \tablenotemark{a }\\
  & & & & & 0.02 $\pm$ 0.21 \tablenotemark{b} & 0.42 $\pm$ 0.12 \tablenotemark{b} \\ 
  & \cite{sou2022} & 5172 $\pm$ 41 & 4.16 $\pm$ 0.10 & -0.04 $\pm$ 0.05 & & \\
  & \cite{2019AJ....158..138S} & 5169 $\pm$ 144 & 4.18 $\pm$ 0.10 & & & \\
  & \cite{2018AJ....155..111L} & 5173 & 4.16 & -0.04 $\pm$ 0.18 & & \\
  & \cite{casa2011} & 5169 $\pm$ 80 & 4.17 & -0.28 & & \\ \hline
\multirow{2}{*}{$\beta$ Pic}  & \multirow{2}{*}{This work} & \multirow{2}{*}{8054 $\pm$ 96} & \multirow{2}{*}{4.18 $\pm$ 0.13} & \multirow{2}{*}{0.01 $\pm$ 0.03} &  -0.25 $\pm$ 0.10\tablenotemark{a} &  0.14 $\pm$ 0.05 \tablenotemark{a}\\ 
  & & & & & -0.27 $\pm$ 0.10\tablenotemark{b} & 0.14 $\pm$ 0.07 \tablenotemark{b} \\ 
  & \cite{2023ApJS..266...11B} & 8102 $\pm$ 81 & 4.10 $\pm$ 0.03 & -0.02 $\pm$ 0.03 & & \\
  & \cite{saffe2021} & 8084 $\pm$ 130 & 4.22 $\pm$ 0.13 & -0.27 $\pm$ 0.14 \tablenotemark{d} & -0.20 $\pm$ 0.11 & \\
  & \cite{swastik2021} & 7890 $\pm$ 17 & 3.83 $\pm$ 0.03 & -0.21 $\pm$ 0.03 \tablenotemark{d} & & \\ 
  & \cite{2019AJ....158..138S} & 8039 & 4.35 & & & \\ \hline
\multirow{5}{*}{YSES 1}  & \multirow{2}{*}{This work} & \multirow{2}{*}{4709 $\pm$ 45} & \multirow{2}{*}{4.28 $\pm$ 0.09} & \multirow{2}{*}{-0.02 $\pm$ 0.03} & -0.04 $\pm$ 0.04 \tablenotemark{a} & 0.05 $\pm$ 0.03 \tablenotemark{a }\\ 
  & & & & & 0.01 $\pm$ 0.14 \tablenotemark{b} & 0.14 $\pm$ 0.14 \tablenotemark{b} \\
  & \CellWithForceBreak{\cite{yses1a} \& \\ \cite{yses2024}} & 4573 $\pm$ 10 & 4.43 $\pm$ 0.02 & -0.07 $\pm$ 0.01 & & \\
  & \cite{2019AJ....158..138S} & 4664 $\pm$ 127 & 4.29 $\pm$ 0.10 & & & \\ \hline
\multirow{2}{*}{51 Eridani} & {This work} & 7175 $\pm$ 75 & 4.24 $\pm$ 0.14 & -0.03 $\pm$ 0.07 & 0.03 $\pm$ 0.07 & 0.15 $\pm$ 0.10 \\
& \cite{2025AJ....169...55B} & 7277 $\pm$ 164 & 4.32 $\pm$ 0.23 & -0.01 $\pm$ 0.11 & 0.03 $\pm$ 0.08 & 0.04 $\pm$ 0.08 \\
\enddata
\tablenotetext{a}{Abundances obtained using spectral fitting}
\tablenotetext{b}{Abundances obtained using equivalent width}
\tablenotetext{c}{Additional literature comparisons made within text (refer Section \ref{subsub:discabpic})}
\tablenotetext{d}{Value corresponds to [Fe/H] and not [M/H]. Difference is significant due to peculiar abundance pattern (refer Section \ref{subsub:discbetapic})}
\end{deluxetable}

\section{Discussion \label{sec:discuss}}

We analyzed a sample of five stars with a wide range of spectral types and measured the abundances of carbon and oxygen using the spectral fitting method and 16 elements:  C, O, Na, Mg, Si, S, K, Ca, Sc, Ti, Cr, Mn, Fe, Ni, Zn, and Y, using the equivalent width method. We obtain abundance measurements for HR 2562, PZ Tel, and YSES 1 for the first time, as well as the first oxygen abundance for $\beta$ Pic. In this work, we obtain the first detailed abundances for HR 2562, PZ Tel, and YSES 1, and the first oxygen measurements for $\beta$ Pic. Additionally, we also utilize our analysis methods for an update of previously measured abundances for AB Pic and $\beta$ Pic, enabling us to have a uniformly analyzed sample of host stars.

\subsection{Comparison with measurements in literature}

\subsubsection{HR 2562}

HR 2562 has had several previous measurements of atmospheric parameters in the literature. The most recent were by \cite{zak2022} and \cite{swastik2021} using archival FEROS and HARPS spectra from the ESO archive\footnote{http://archive.eso.org/scienceportal/}. All their parameters agree with ours within a maximum of 2.5$\sigma$. Additionally, all our parameters agree very well with those reported by \cite{mesa2018} who used a combination of UVES and FEROS spectra to determine atmospheric properties of the HR 2562 primary. Lastly, our findings are also in agreement with those published by \cite{casa2011} in their reanalysis of parameters in the Geneva-Copenhagen Survey \citep{2004A&A...418..989N}. 

\subsubsection{AB Pic \label{subsub:discabpic}}

As a slower-rotating K1V star $\sim$50 pc from the Sun \citep{gaiaedr3}, AB Pic has been extensively studied in the literature. Perhaps the most directly comparable previous study is from the APOGEE-2 data release \citep{2022ApJS..259...35A}, which measured the stellar properties and several elemental abundances. Of the stellar properties, $T_{\mathrm{eff}}$ deviates the most ($\sim$3$\sigma$), while $\log{g}$ and [M/H] agree within 1$\sigma$. For elemental abundances, [C/H] agrees within $1.5\sigma$ while [O/H] agrees within 2$\sigma$. Another study is by \cite{2018AJ....155..111L} who also obtain carbon and oxygen abundances. While their stellar parameters agree with ours within $2\sigma$, the lack of uncertainties on their elemental abundances make comparisons difficult. Summarizing the results of the remaining literature on this target (e.g., \citealt{2010ApJ...720.1290G, casa2011, 2014MNRAS.442..220F, 2018A&A...614A..55A, 2019AJ....158..138S,swastik2021, sou2022}), we find values $T_{\mathrm{eff}}$ = 5027--5378 K, $\log{g}$ = 4.44--4.63, [M/H] $= -0.05$--0.09. Our stellar atmospheric parameters fall well within this range.

\subsubsection{PZ Tel}

Even though it is late-G star $<$ 50 pc from the Sun \citep{gaiaedr3}, PZ Tel does not have previous elemental abundances due to its high rotational velocity ($v\sin{i}$ $\sim$ 70 kms$^{-1}$). Existing literature includes measurements of stellar atmospheric parameters by \cite{casa2011}, \cite{2018AJ....155..111L}, \cite{2019AJ....158..138S}, and \cite{sou2022} (Table \ref{tab:litcompare}). Our determined parameters agree with all of those to $\lesssim$2$\sigma$.

\subsubsection{\texorpdfstring{$\beta$}{Beta} Pic \label{subsub:discbetapic}}

Four previous studies have determined stellar parameters for $\beta$ Pic (Table \ref{tab:litcompare}).  Our $T_{\mathrm{eff}}$ and $\log{g}$ are in 1$\sigma$ agreement with all of them except the $\log{g}$ by \cite{swastik2021}, which agrees only at 2.5$\sigma$. However, there seems to be considerable variation in the [M/H] values for $\beta$ Pic, with our work and \cite{2023ApJS..266...11B} getting solar metallicities, and \cite{saffe2021} and \cite{swastik2021} getting metallicities 0.2 dex below solar values. This is easily resolved when considering the peculiar abundance profile of $\beta$ Pic. As seen in Table \ref{tab:elemntandratio}, $\beta$ Pic has sub-solar abundances of Mg, Si, Sc, Cr, Mn, Fe, and Ni, and solar/super-solar abundances of Na, S, K, Ca, Ti, Zn, and Y. This implies that the [M/H] value obtained depends on dominant species in the spectral regions used for metallicity determination. If we consider the iron abundance, our [Fe/H] $= -0.23 \pm 0.20$ agrees very well with the [Fe/H] values from \cite{saffe2021} and \cite{swastik2021}. 

\par \cite{saffe2021} previously determined abundances of C I, Mg I, Mg II, Al I, Si II, Ca I, Ca II, Sc II, Ti II, Cr II, Mn I, Fe I, Fe II, Sr II, Y II, Zr II, and Ba II for $\beta$ Pic. The comparisons are outlined in Table \ref{tab:betapic}, our abundances agree excellently with their values (within 1$\sigma$).

\subsubsection{YSES 1}

The stellar properties of YSES 1 have been determined as part of studies on the properties of its two companions \citep{yses1a, yses2024} and in the TESS input catalog \citep{2019AJ....158..138S}. Our $T_{\mathrm{eff}}$ deviates from the former significantly ($\sim$ 3$\sigma$) but agrees with the latter values. However our $\log{g}$ and [M/H] agree with both literature estimates within $\sim$1.5$\sigma$.

\begin{deluxetable}{ccc}
\tabletypesize{\small} 
\tablecaption{Abundance comparisons of this work with \cite{saffe2021} for $\beta$ Pic \label{tab:betapic}}
\tablewidth{0pt}
\tablehead{
\colhead{Species} & \colhead{This work} & \colhead{\cite{saffe2021}}}
\startdata
C \textsc{i}   & $-0.27 \pm 0.10$ & $-0.20 \pm 0.11 $ \\
Mg \textsc{i}  & $-0.23 \pm 0.14$ & $-0.22 \pm 0.20 $\\
Si \textsc{ii} & $-0.18 \pm 0.05$ \tablenotemark{*}  & $-0.14 \pm 0.24 $ \\
Ca \textsc{ii}   & $-0.03 \pm 0.20$ & $0.11 \pm 0.38 $ \\
Sc \textsc{ii} & $-0.17 \pm 0.05$ & $-0.14 \pm 0.25 $ \\
Ti \textsc{ii}  & $-0.04 \pm 0.10$ & $-0.08 \pm 0.19 $ \\
Cr \textsc{ii}  & $-0.17 \pm 0.05$ & $-0.18 \pm 0.10 $ \\
Mn \textsc{ii}   & $-0.31 \pm 0.11$ & $-0.23 \pm 0.13 $ \\
Fe \textsc{ii}   & $-0.24 \pm 0.20$ & $-0.28 \pm 0.14 $ \\
Fe \textsc{ii}  & $-0.21 \pm 0.19$ & $-0.26 \pm 0.14 $ \\
Y \textsc{ii}   &  $0.03 \pm 0.09$ & $-0.01 \pm 0.09 $ \\
\enddata
\tablenotetext{*}{Si I used instead of Si II due to unresolved blend in Si II.}
\end{deluxetable}

\subsection{Elemental abundance trends among directly imaged planet host stars}

\begin{figure}
    \centering
    \includegraphics[width=0.9\linewidth]{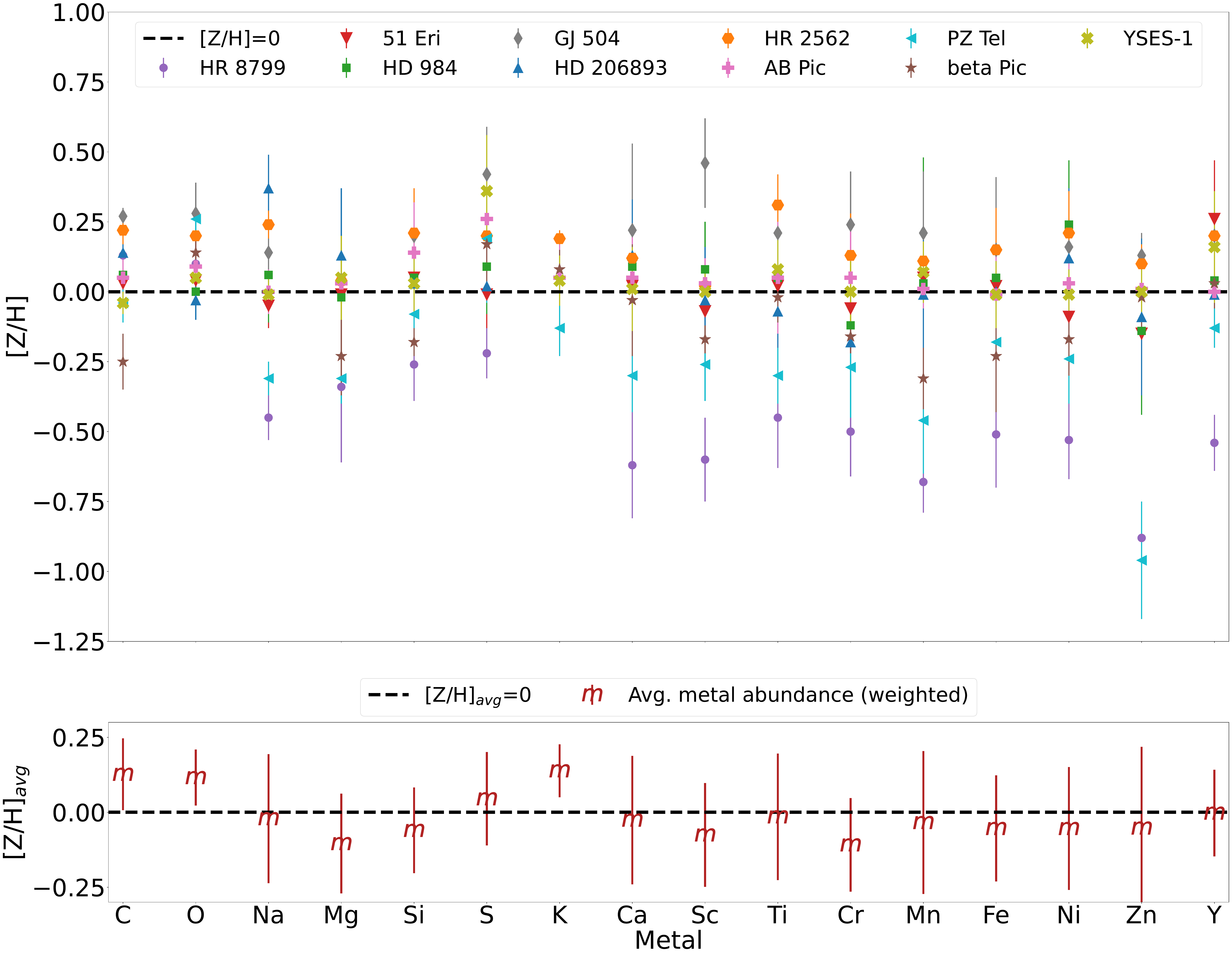}
    \caption{(\textit{Top}) Abundances relative to solar for the five targets in this work (from Table \ref{tab:ewabundance}) and the host stars HR 8799, 51 Eri, HD 984, GJ 504, and HD 206893 from Paper I. As we did not measure the potassium abundance in Paper I, we do not report its abundance for those five stars.
    (\textit{Bottom}) Population-level average abundance of the 16 elements relative to solar values. Carbon and oxygen have super-solar abundances at the population level, while all other elements are solar. We do not consider potassium as its abundance is measured only for the five targets in this work.}
    \label{fig:elementplot}
\end{figure}

We use the abundance measurements for the spectral species in Table \ref{tab:elemntandratio} to calculate elemental abundances relative to solar for all targets, looking for possible trends in elemental abundances for directly imaged companion host stars. In addition, we use the carbon and oxygen abundances determined using both spectral fitting and the equivalent width method to calculate C/O ratios for the targets. Equivalent width abundances of carbon, oxygen, and sulfur are used to compute the C/S and O/S ratio. The various abundances and abundance ratios are compiled in Table \ref{tab:elemntandratio}. Combining these results with those from Paper I gives us a sample of 10 directly imaged companion host stars with detailed abundances. We plot the elemental abundances for all 10 targets in Figure \ref{fig:elementplot} (\textit{Top}). 

\par HR 2562 shows enhanced abundances for all elements, corresponding to its super-solar overall metallicity. The sulfur abundance for the host stars AB Pic, PZ Tel, and YSES 1 (all late-G and K-type stars) have large measurement error due to the weak spectral lines, hence we do not draw conclusions with regard to the observed super-solar sulfur abundance. Ignoring sulfur, PZ Tel shows sub-solar abundances for all elements except carbon and oxygen. Carbon is solar and oxygen is super-solar. We also observe the peculiar abundance pattern for $\beta$ Pic as discussed previously in section \ref{subsub:discbetapic}, with solar or super-solar O, Na, S, K, Ca, Ti, Zn, and Y and sub-solar C, Mg, Si, Sc, Cr, Mn, Fe, and Ni. For the targets from Paper I, we observe super-solar abundances for GJ 504 and sub-solar elemental abundances for HR 8799 (excepting C and O), corresponding to its $\lambda$ Bootis nature \citep{1999AJ....118.2993G}. 

\par In order to investigate any population-level deviations from solar values for these 16 elements for our sample of directly imaged host stars, we calculated the sample average for each elemental abundances ([Z/H]$_{avg}$). For each target, the measured abundance of an element is weighted using the inverse of the square of the standard deviation ($\sigma_{[Z/H]}^2$). The population-level abundances are plotted in Figure \ref{fig:elementplot} (\textit{Bottom}). Carbon, oxygen, and potassium show super-solar abundances to 1-sigma, while all others are solar. As there are no potassium abundances for the targets from Paper I, we ignore the apparent super solar abundance for now. 

\par However, the super-solar carbon and oxygen merit greater discussion. Apart from hydrogen, helium, and the noble gases, carbon, oxygen, and nitrogen are the only elements present in the volatile phase (gas and ices) in the protoplanetary disk and play a significant role in planet formation (e.g., \citealt{2014A&A...570A..35M,2016A&A...595A..83E,turrini2021}). Both pebble \citep{2017AREPS..45..359J} and gas accretion \citep{2021MNRAS.501.2017N} play a major role in the growth of giant planets. However, once the pebble isolation mass is reached, only gas accretion occurs, with a higher efficiency of gas accretion (e.g., due to lower disk viscosity) leading to more massive planets \citep{sb2021b}. These factors make the volatile species likely more important for gas giant formation. Additionally, for the wide orbit directly imaged planets, a larger ice content in the disk at their formation location could lead to a decrease in the critical core mass required for gas giant formation (e.g., \citealt{2011MNRAS.416.1419H, 2014A&A...572A..35L, 2015A&A...576A.114V}). Future measurements of the nitrogen abundance for these host stars are needed to investigate this intriguing potential abundance trend.

\subsection{Comparison of stellar and companion C/O ratios}

As discussed earlier, the C/O ratio is an important diagnostic for constraining planet formation and has been measured for many directly imaged substellar companions. Gravitational instability is expected to lead to planetary C/O ratios similar to that of the host star. For planet formation by core/pebble accretion, the formation location relative to the snowlines of H$_2$O, CO$_2$, and CO could lead to a range of C/O ratios \citep{oberg2011,piso2015}. Among the stars in this work, we find solar C/O ratios for three
targets (HR 2562, AB Pic, and YSES 1), and sub-solar C/O ratios at a $4\sigma$ significance for PZ Tel and $\beta$ Pic. Together with the five targets in Paper I, we have C/O ratio measurements for 10 directly imaged companion host stars. However, not all companions around these targets have estimates of C/O ratio, with measurements only available for AB Pic b, $\beta$ Pic b, YSES 1 b, and YSES 1 c (Table \ref{tab:companiontable}). 

\par
\begin{figure}
    \centering
    \includegraphics[width=0.8\linewidth]{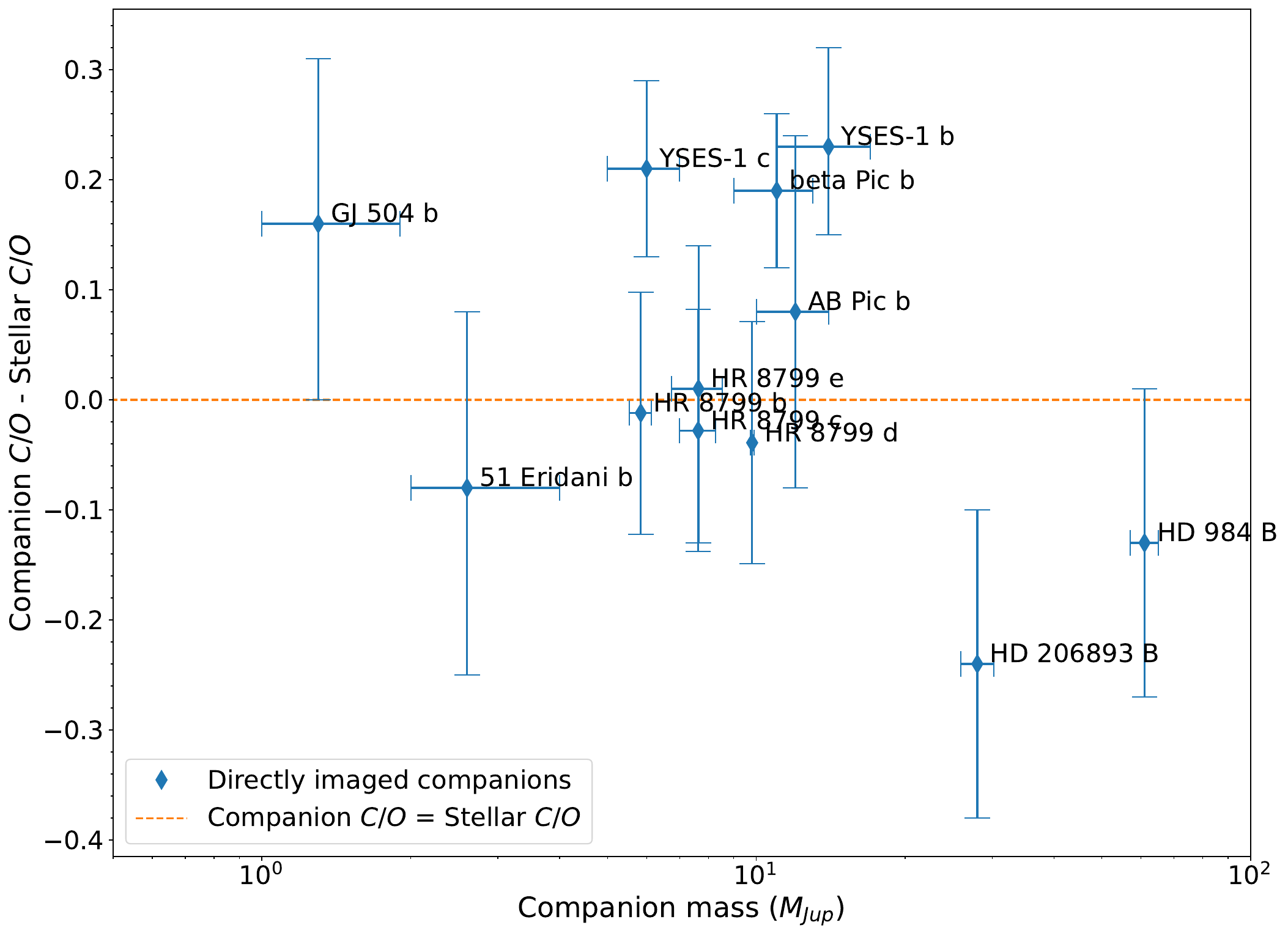}
    \caption{Comparison of C/O ratios for stellar targets from Paper I and this work to those of their substellar companions. The dashed line (orange) denotes companion C/O ratio being equal to the stellar C/O ratio. All values are encapsulated in Table \ref{tab:companiontable}}
    \label{fig:cocompare}
\end{figure}

 In Figure \ref{fig:cocompare}, we compare companions with available C/O ratios to those of their host star determined in Paper I or in this work. GJ 504 b was found to have super-stellar metallicity \citep{malin2025} and its super-stellar C/O imply planet-like formation. \cite{hoch2025} measured super-stellar metallicities for YSES 1 b and c, which in conjunction with their super-stellar C/O ratios favor formation by core accretion and subsequent migration to their current positions. AB Pic b also presents a similar case to the YSES 1 planets with its wide orbit and slightly super-stellar metallicity \citep{palma2023}; however, its C/O is stellar. 

$\beta$ Pic b has a super-stellar C/O ratios, which along with its debris disk ($\beta$ Pic system, e.g., \citealt{1984Sci...226.1421S, 2003ApJ...584L..27W}) and possibly super-stellar metallicities \citep{gravity2020, 2024AJ....167...45R} suggest its formation by core accretion-type mechanisms. However, the peculiar chemical profile of $\beta$ Pic suggest that using its abundance profile might not be a suitable proxy for the protoplanetary disk from which $\beta$ Pic b and c formed \citep{2024AJ....167...45R}. If we instead use the abundances of HD 181327 (C/O = 0.62 $\pm$ 0.08; \citealt{2024AJ....167...45R}), an F dwarf in the $\beta$ Pic moving group as a proxy for the abundances in the protoplanetary disk of $\beta$ Pic, we get a sub-stellar C/O ratio for the planet. This scenario suggests accretion of oxygen-rich ices after initial formation by core-accretion between the H$_2$O and CO$_2$ ice lines.

\par Among the remaining companions, the HR 8799 planets and 51 Eridani b all have stellar C/O, but the large uncertainties prevent drawing definitive conclusions. The high mass of HD 984 B ($\sim$61 $M_\mathrm{Jup}$; \citealt{2022AJ....163...50F}) and stellar C/O indicates a strong possibility of gravitational instability-like formation. While HD 206893 B appears to have a sub-stellar C/O, it agrees with stellar values when using the equivalent width C/O from Paper I (0.69 $\pm$ 0.35). The presence of a debris disk in the system strongly suggest formation in a disk and post-formation solid accretion from the disk, leading to sub-stellar C/O ratios \citep{turrini2021}. 

\begin{deluxetable}{cccccc}
\tabletypesize{\small}
\tablecaption{Companion properties and stellar C/O ratios\label{tab:companiontable}}
\tablewidth{0pt}
\tablehead{\colhead{Companion} & \colhead{Mass ($M_\mathrm{Jup}$)}    & \CellWithForceBreak{Companion \\ C/O Ratio} & \CellWithForceBreak{Semi-Major Axis \\ (au)} & \CellWithForceBreak{Stellar \\ C/O Ratio \tablenotemark{*}} & \colhead{References}} 
\startdata
\multirow{2}{*}{51 Eri b} & $\leq$11 (Dynamical)   & \multirow{2}{*}{$0.46^{+0.08}_{-0.10}$}  & \multirow{2}{*}{$11.1^{+4.2}_{-1.3}$} & \multirow{2}{*}{0.54 $\pm$ 0.14} & \multirow{2}{*}{1,2,3,4,5,6,7,8,39}\\
& 2--9 (Evolutionary) & & & & \\ \hline
HR 8799 b &   5.84 $\pm$ 0.3   & $0.578^{+0.004}_{-0.005}$ & 67.96 $\pm$ 1.85 & \multirow{4}{*}{0.59 $\pm$ 0.11} & {9,10,11,36,39} \\
HR 8799 c &  $7.63^{+0.64}_{-0.63}$ & 0.562 $\pm$ 0.004       & 42.81 $\pm$ 1.16 & & {9,10,11,36,39} \\
HR 8799 d &  9.81 $\pm$ 0.08  & $0.551^{+0.005}_{-0.004}$     & 26.97 $\pm$ 0.73 & & {9,10,11,36,39} \\
HR 8799 e &  $7.64^{+0.89}_{-0.91}$ & $0.60^{+0.07}_{-0.08}$  & 16.99 $\pm$ 0.46 & & {9,10,12,36,39} \\
\hline
HD 984 B & 61 $\pm$ 4   &  0.50 $\pm$ 0.01 & 28$^{+7}_{-4}$ & 0.63 $\pm$ 0.14 & {13,17,39} \\ 
\hline
\multirow{2}{*}{GJ 504 b} & $1.3^{+0.6}_{-0.3}$ (Young system) &  \multirow{2}{*}{$0.70^{+0.06}_{-0.07}$} & \multirow{2}{*}{$\sim$ 43.5} & \multirow{2}{*}{0.54 $\pm$ 0.14} & \multirow{2}{*}{14,34,37,39} \\
& $23.0^{+10}_{-9}$ (Old system) & & & &  \\ 
\hline
HD 206893 B & $28.0^{+2.2}_{-2.1}$  & 0.57 $\pm$ 0.02 & $8.93^{+1.41}_{-0.19}$ & \multirow{2}{*}{0.81 $\pm$ 0.14} & {15,16,39} \\
HD 206893 c & $12.7^{+1.2}_{-1.0}$  & & $3.62^{+0.35}_{-0.42}$ & & {15,16,39} \\ 
\hline
\multirow{2}{*}{HR 2562 B} & $\leq$18.5 (Dynamical) & & \multirow{2}{*}{$30^{+11}_{-8}$} & \multirow{2}{*}{0.58 $\pm$ 0.09} & \multirow{2}{*}{18,19,35,38,40}\\
& 14 $\pm$ 2 (Evolutionary) & & & & \\ 
\hline
AB Pic b & 12 $\pm$ 2 & 0.58 $\pm$ 0.08 & 190--260 & 0.50 $\pm$ 0.14 & {20,21,22,23,24,40}\\ 
\hline
PZ Tel B & $27^{+25}_{-9}$ & & $27^{+14}_{-4}$ & 0.28 $\pm$ 0.05 & {25,40} \\ 
\hline
$\beta$ Pic b & 11 $\pm$ 2 & 0.41 $\pm$ 0.04 & 9.5 $\pm$ 0.5 & \multirow{2}{*}{0.22 $\pm$ 0.06}& {26,27,28,29,30,40} \\
$\beta$ Pic c & 10 $\pm$ 1 & & 2.68 $\pm$ 0.02 & & {28,29,40} \\ 
\hline
YSES 1 b & 14 $\pm$ 3 & 0.62--0.75 & $\sim$ 162 & \multirow{2}{*}{0.45 $\pm$ 0.05} & {31,33,40}\\
YSES 1 c & 6 $\pm$ 1 & 0.60--0.72 & $\sim$ 320 & & {32,33,40} \\
\enddata
\tablerefs{(1) \cite{2020AJ....159....1D}, (2) \cite{2022MNRAS.509.4411D}, (3) \cite{doi:10.1126/science.aac5891}, (4) \citet{doi:10.1051/0004-6361/201629767}, (5) \cite{2019AJ....158...13N}, (6) \cite{2025arXiv251008327M}, (7) \cite{2023MNRAS.525.1375W}, (8) \cite{2024PASA...41...43E}, (9) \cite{2022AJ....163...52S}, (10) \cite{doi:10.1051/0004-6361/202243862}, (11) \cite{2021AJ....162..290R}, (12) \cite{doi:10.1051/0004-6361/202038325}, (13) \cite{2022AJ....163...50F}, (14) \cite{bonnefoy2018}, (15) \cite{doi:10.1051/0004-6361/202244727}, (16) \cite{2025AJ....169..175S}, (17) \cite{2024AA...686A.294C}, (18) \cite{2023AJ....165..219Z}, (19) \cite{godoy2024}, (20) \cite{chauvin2005}, (21) \cite{bonnefoy2010}, (22) \cite{bonnefoy2014}, (23) \cite{palma2023}, (24) \cite{2025MNRAS.537..134G}, (25) \cite{2023AJ....165..246F}, (26) \cite{2016AJ....152...97W}, (27) \cite{2018NatAs...2..883S}, (28) \cite{2019NatAs...3.1135L}, (29) \cite{2022ApJS..262...21F}, (30) \cite{landman2024}, (31) \cite{yses1a}, (32) \cite{yses1b}, (33) \cite{hoch2025}, (34) \cite{malin2025}, (35) \cite{2016ApJ...829L...4K}, (36) \cite{zurlo2016}, (37) \cite{2013ApJ...774...11K}, (38) \cite{maire2018}, (39) \cite{2025AJ....169...55B}, (40) This work }
\tablenotetext{*}{Spectral fit C/O ratios used}
\end{deluxetable}

\subsection{Directly imaged host star C/O relative to transit and radial velocity planet hosts}

While the younger directly imaged planet host stars are poorly studied, extensive abundance measurements have been made for the F, G, K-type host stars of planets discovered using the transit and radial velocity (RV) methods. The radial velocity planet host stars have abundances for stars included in the HARPS sample (e.g., \citealt{2015A&A...576A..89B, 2017A&A...599A..96S}), while for the transiting planets there exist smaller surveys for specific planet populations (e.g., \citealt{2014ApJ...788...39T} for transiting hot Jupiters) as well as the larger APOGEE-KOI Goal program in APOGEE DR16 \citep{2020ApJS..249....3A,2020AJ....160..120J}. 

\par
\begin{figure}
    \centering
    \includegraphics[width=0.8\linewidth]{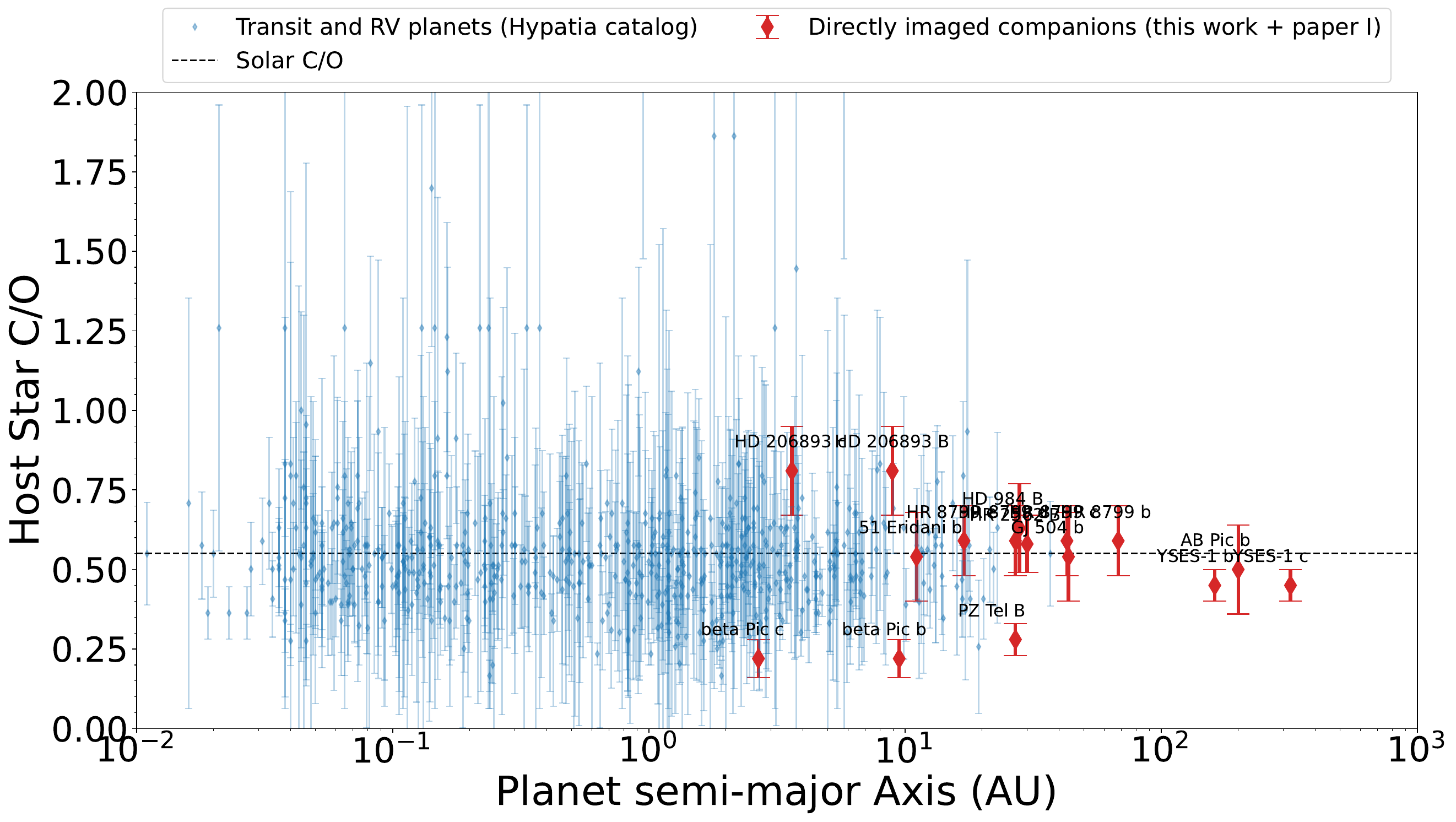}
    \caption{Host star C/O ratios for transiting and RV planets (blue) versus directly imaged companion host stars (red). The dashed black line indicates the solar C/O ratio. The detection bias involved in direct imaging is evident with the higher semi-major axes of the the directly imaged companions. We do not notice any significant differences between our directly imaged host stars and the transit + RV host star populations, which is corroborated by our KS Test.}
    \label{fig:ditransitrv}
\end{figure}

 In order to investigate the chemical profile of the directly imaged companion host star population relative to the host stars with planets discovered through transit and/or radial velocity, we plot the C/O of the two stellar populations together in Figure \ref{fig:ditransitrv}. We plot the semi-major axis of the planet/companion on the x-axis to clearly highlight the two populations being compared. The semi-major axis and host star C/O for the wide orbit population is from Table \ref{tab:companiontable}, while the corresponding parameters for other planetary populations have been adopted from the Hypatia Catalog \citep{hinkel2014}.

\par
\begin{figure}
    \centering
    \includegraphics[width=0.4\linewidth]{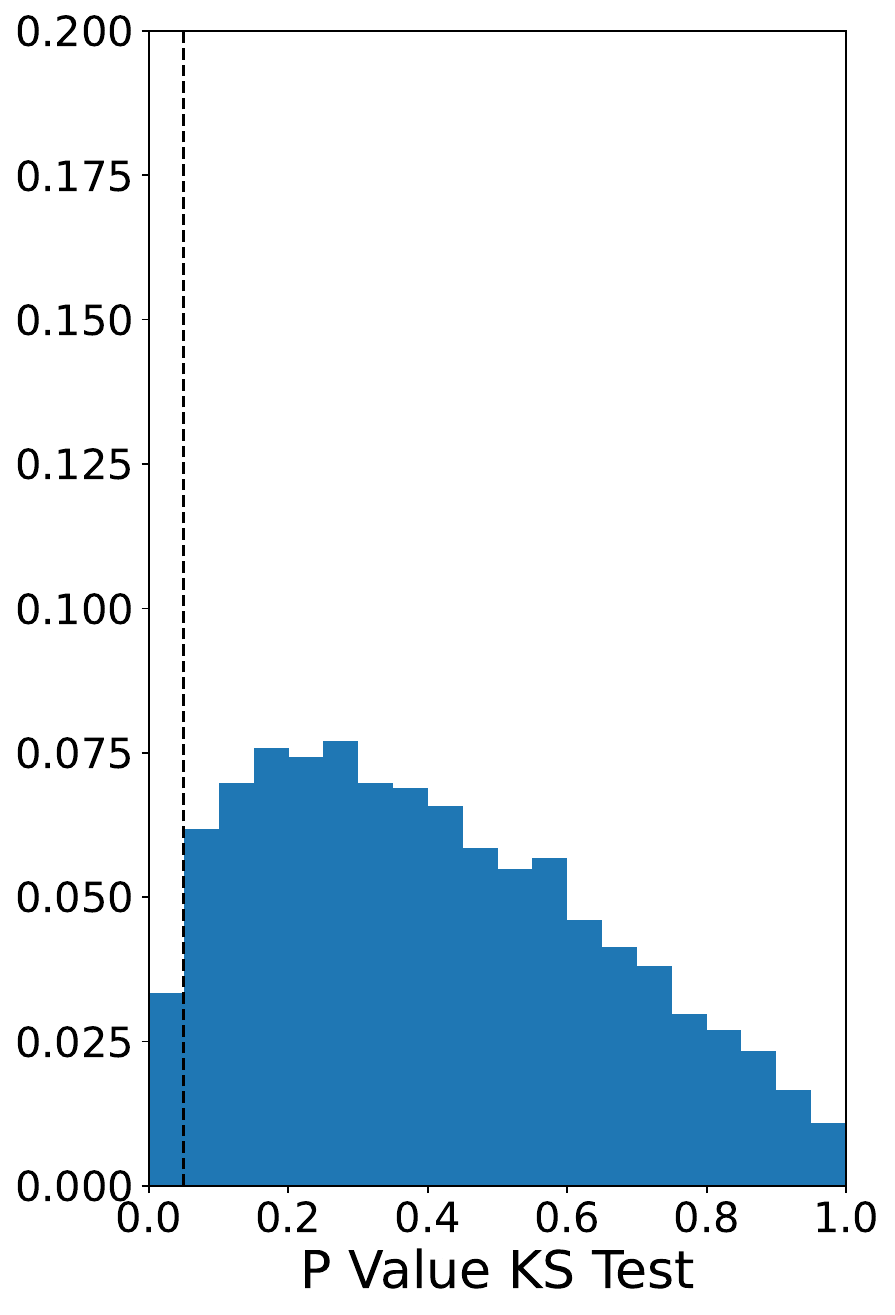}
    \caption{A histogram of p-values from the KS Test. The dashed black line indicates a p-value of 0.05. We obtain p $>$ 0.05 for 96.7\% of the 100,000 iterations, \textbf{indicating that we do not have enough evidence to conclude that both stellar sub-samples are from different populations.}}
    \label{fig:kstest}
\end{figure}

 We perform the Kolmogorov–Smirnov (KS) test to determine if the two samples come from the same parent distribution. For each star in both populations, we randomly sample a C/O value from a Gaussian distribution centered on the central C/O value with the uncertainty as the standard deviation. The KS test is performed on the randomly sampled C/O ratios for the two sub-samples using the \texttt{scipy.stats.ks$\_$2samp} Python package \citep{2020SciPy-NMeth}. After repeating the procedure 100,000 times, a histogram of KS p-values is obtained (Figure \ref{fig:kstest}). We obtain p-value $>$ 0.05 for 96,700 of the 100,000 iterations (96.7\%), \textbf{indicating that we do not have enough evidence to conclude that the sub-samples come from different parent distributions}. In addition, we also observe that the precisions on our measurements are higher ($<$0.15 dex) than those of other sources in literature (obtained using a variety of instruments and resolutions), showcasing the potential of our analysis procedures to obtain high-precision abundances, even for the young and fast-rotating host stars.

\section{Conclusions \label{sec:conclusion}}


In this paper, we build upon our previous work measuring detailed elemental abundances for fast-rotating F/G-type directly imaged host stars (Paper I; \citealt{2025AJ....169...55B}). We utilize the high-resolution ($R$ $\sim$ 76,000) GHOST spectrograph at Gemini-South to obtain high signal-to-noise (SNR $\geq$ 200) spectra of five host stars with companions of interest to the exoplanet community: HR 2562, AB Pic, PZ Tel, $\beta$ Pic, and YSES 1. These host stars also showcase the applicability of techniques developed in paper I to K and A-type stars, as well as to targets with $v\sin{i}$ $\sim$ 120 kms$^{-1}$. Among these targets, only AB Pic and $\beta$ Pic have some elemental abundances in the literature, with none of them having the full suite of abundance estimates crucial to planet formation studies (C, O, Na, Mg, Si, S, K, Fe, Ni). 

\par We use the spectral fit method to measure the abundances of C and O for all five targets, and hence the C/O ratio. We find a super-solar oxygen for PZ Tel ($>$ 5$\sigma$) and $\beta$ Pic ($>$ 2$\sigma$), and a sub-solar carbon for $\beta$ Pic ($>$ 2$\sigma$) in line with previous estimations by \cite{saffe2021}. Our results indicate diversity in stellar C/O ratios, with HR 2562, AB Pic, and YSES 1 having solar C/O ratios, and PZ Tel and $\beta$ Pic having sub-solar C/O ratios at a $4\sigma$ significance. While the C/O uncertainties are higher for the equivalent width method, our findings hold up at similar significance levels, further validating the robustness of our results. Our findings indicate that assuming solar abundances and C/O ratios for host stars of wide orbit companions will lead to inaccurate conclusions regarding their formation.

\par In addition, we utilize equivalent widths to measure the abundances of Na, Mg, Si, S, K, Ca, Sc, Ti, Cr, Mn, Fe, Ni, Zn, and Y, with our exceptional data quality enabling us to obtain precisions $<$0.2 dex on all our abundances, even for the extremely fast-rotating $\beta$ Pic ($v\sin{i}$ $\sim$ 120 kms$^{-1}$). Among these 14 elements, we find super-solar abundances for HR 2562, sub-solar abundances for PZ Tel, and solar abundances for AB Pic and YSES 1. We also confirm the unusual abundance profile of $\beta$ Pic, with its sub-solar C, Mg, Si, Sc, Cr, Mn, Fe, and Ni, super-solar O and potentially S, and solar Na, K, Ca, Ti, Zn, and Y.

\par Subsequently, we use the host star abundances from this work and Paper I to investigate potential abundance trends among the wide orbit planet host star population using a weighted average abundance for each element. We notice super-solar abundances of carbon and oxygen, with all other elements being solar, indicating that stars hosting wide-orbit gas giants might potentially be rich in volatile elements. These findings line up well with theoretical models indicating that accretion of icy (volatile-rich) planetesimals during the core accretion phase might reduce the critical core mass required for gas giant formation. Future stellar abundance measurements for nitrogen would enable further investigation and validate any potential trends.

\par We also compare the stellar C/O ratios to those of their companions with available C/O measurements. The super-stellar C/O for GJ 504 b, $\beta$ Pic b, AB Pic b, YSES 1 b, and YSES 1 c, along with their super-stellar metallicities in the literature, strongly indicate a planet-like formation for these companions. However, the peculiar chemical abundances of $\beta$ Pic invite caution while using its abundances to uncover the formation of $\beta$ Pic b and c. The system architecture strongly indicates a planet-like formation for HD 206893 B, with potential post-formation solid accretion leading to a sub-stellar C/O. The high mass of HD 984 B suggests formation through gravitational instability or a binary-like mechanism. Lastly, the large error bars on the HR 8799 planets and 51 Eridani preclude any conclusions regarding their formation mechanism. While observations of these wide-orbit companions using ground-based instruments (e.g., CRIRES+ on VLT) and JWST are enabling more accurate estimates of planetary metallicities and C/O ratios, the degeneracies involved in using the C/O ratios as a formation imprint make it imperative to consider using volatile-to-refractory and other elemental ratios (C/S, O/S, etc.) as additional formation tracers. 

\par Lastly, we compare the C/O of our directly imaged host star sample with the transiting and radial velocity planet host stars. On performing 100,000 randomly-sampled KS tests, we obtain p-value $>$ 0.05 for $\sim$98\% of the iterations, \textbf{indicating a lack of sufficient evidence to conclude that the two host star samples come from different populations}.

\begin{deluxetable}{cccccccccccc}
\rotate
\tablecaption{Elemental abundances (relative to solar) and abundance ratios for targets in this paper \label{tab:elemntandratio}}
\tablewidth{0pt}
\tablehead{
\colhead{Element} & \colhead{HR 2562} & \colhead{} & \colhead{AB Pic} & \colhead{} & \colhead{PZ Tel} & \colhead{} & \colhead{$\beta$ Pic} & \colhead{} & \colhead{YSES 1} & \colhead{} \\
\colhead{(Z)} & \colhead{[Z/H]} & \colhead{$\sigma_{[Z/H]}$} & \colhead{[Z/H]} & \colhead{$\sigma_{[Z/H]}$} & \colhead{[Z/H]} & \colhead{$\sigma_{[Z/H]}$} & \colhead{[Z/H]} & \colhead{$\sigma_{[Z/H]}$} & \colhead{[Z/H]} & \colhead{$\sigma_{[Z/H]}$}
}
\startdata
C\tablenotemark{*}  & 0.22 & 0.05 & 0.05 & 0.10 & -0.04 & 0.07 & -0.25 & 0.10 & -0.04 & 0.04 \\
O\tablenotemark{*}  & 0.20 & 0.05 & 0.09  & 0.07 & 0.26  & 0.04 & 0.14  & 0.05 & 0.05  & 0.03 \\
Na & 0.24 & 0.05 & 0.00  & 0.08 & -0.31 & 0.06 & -0.01 & 0.05 & -0.01 & 0.06 \\
Mg & 0.04 & 0.13 & 0.03  & 0.11 & -0.31 & 0.09 & -0.23 & 0.14 & 0.05  & 0.15 \\
Si & 0.21 & 0.16 & 0.14  & 0.18 & -0.08 & 0.09 & -0.18 & 0.05 & 0.03  & 0.12 \\
S  & 0.20 & 0.13 & 0.26  & 0.19 & 0.19  & 0.23 & 0.17  & 0.14 & 0.36  & 0.20 \\
K  & 0.19 & 0.03 & 0.05  & 0.12 & -0.13 & 0.10 & 0.08  & 0.07 & 0.04  & 0.09 \\
Ca & 0.12 & 0.09 & 0.05  & 0.15 & -0.30 & 0.13 & -0.03 & 0.20 & 0.01  & 0.15 \\
Sc & 0.02 & 0.07 & 0.03  & 0.09 & -0.26 & 0.13 & -0.17 & 0.05 & 0.00  & 0.08 \\
Ti & 0.31 & 0.11 & 0.05  & 0.20 & -0.30 & 0.10 & -0.02 & 0.09 & 0.08  & 0.11 \\
Cr & 0.13 & 0.15 & 0.05  & 0.18 & -0.27 & 0.18 & -0.16 & 0.06 & 0.00  & 0.11 \\
Mn & 0.11 & 0.05 & 0.01  & 0.07 & -0.46 & 0.19 & -0.31 & 0.11 & 0.07  & 0.11 \\
Fe & 0.15 & 0.15 & -0.01 & 0.17 & -0.18 & 0.16 & -0.23 & 0.20 & -0.01 & 0.12 \\
Ni & 0.21 & 0.15 & 0.03  & 0.10 & -0.24 & 0.16 & -0.17 & 0.13 & -0.01 & 0.09 \\
Zn & 0.10 & 0.07 & 0.01  & 0.04 & -0.96 & 0.21 & -0.02 & 0.15 & 0.00  & 0.08 \\
Y  & 0.20 & 0.08 & 0.00  & 0.05 & -0.13 & 0.07 & 0.03  & 0.09 & 0.16  & 0.20 \\
\hline\hline
\colhead{Abundance ratio} & \colhead{HR 2562} & \colhead{} & \colhead{AB Pic} & \colhead{} & \colhead{PZ Tel} & \colhead{} & \colhead{$\beta$ Pic} & \colhead{} & \colhead{YSES 1} & \colhead{} & \colhead{Solar} \\
\colhead{} & \colhead{X/Y} & \colhead{$\Delta$X/Y} & \colhead{X/Y} & \colhead{$\Delta$X/Y} & \colhead{X/Y} & \colhead{$\Delta$X/Y} & \colhead{X/Y} & \colhead{$\Delta$X/Y} & \colhead{X/Y} & \colhead{$\Delta$X/Y} & \\
\hline
C/O (spectral fit) & 0.58  & 0.09 & 0.50  & 0.14  & 0.28  & 0.05  & 0.22  & 0.06  & 0.45  & 0.05  & \multirow{2}{*}{0.55}  \\
C/O (eq. width)    & 0.59  & 0.23 & 0.37  & 0.20  & 0.22  & 0.12  & 0.21  & 0.06  & 0.41  & 0.19  &       \\
C/S                & 15.14 & 6.66 & 12.59 & 6.67  & 13.80 & 9.90  & 7.41  & 2.94  & 9.12  & 5.13  & 20.42 \\
O/S                & 25.70 & 9.71 & 33.88 & 21.52 & 63.10 & 37.69 & 34.67 & 12.50 & 22.39 & 12.55 & 37.15
\enddata
\tablenotetext{*}{Due to their high precision, we use the spectral fit abundance values for C and O.}
\end{deluxetable}

\section*{acknowledgements}
The authors would like to thank Henrique Reggiani for his support in obtaining these observations. A.B. would like to thank Anish Amarsi for the helpful discussions regarding NLTE effects for oxygen. A.B. and Q.M.K acknowledge support by the National Aeronautics and Space Administration under Grants/Contracts/Agreements No. 80NSSC24K0210 issued through the Astrophysics Division of the Science Mission Directorate.
This research has made use of the NASA Exoplanet Archive, which is operated by the California Institute
of Technology, under contract with the National Aeronautics and Space Administration under the Exoplanet Exploration Program.
Any opinions, findings, conclusions, and/or recommendations expressed in this paper are those of the author(s) and do not reflect the views of the National Aeronautics and Space Administration. This work is based on observations obtained at the international Gemini Observatory, a program of NSF NOIRLab, which is managed by the Association of Universities for Research in Astronomy (AURA) under a cooperative agreement with the U.S. National Science Foundation on behalf of the Gemini Observatory partnership: the U.S. National Science Foundation (United States), National Research Council (Canada), Agencia Nacional de Investigaci\'{o}n y Desarrollo (Chile), Ministerio de Ciencia, Tecnolog\'{i}a e Innovaci\'{o}n (Argentina), Minist\'{e}rio da Ci\^{e}ncia, Tecnologia, Inova\c{c}\~{o}es e Comunica\c{c}\~{o}es (Brazil), and Korea Astronomy and Space Science Institute (Republic of Korea). The data were obtained under Program IDs GS-2024B-Q-311 and GS-2025A-Q-306. This work used Bridges-2 \citep{osti_10299128} at Pittsburgh Supercomputing Center through allocation PHY230140 from the Advanced Cyberinfrastructure Coordination Ecosystem: Services \& Support (ACCESS) program, which is supported by National Science Foundation grants \#2138259, \#2138286, \#2138307, \#2137603, and \#2138296. This work has made use of the VALD database, operated at Uppsala University, the Institute of Astronomy RAS in Moscow, and the University of Vienna. The research shown here acknowledges use of the Hypatia Catalog Database, an online compilation of stellar abundance data as described in Hinkel et al. (2014, AJ, 148, 54), which was supported by NASA's Nexus for Exoplanet System Science (NExSS) research coordination network and the Vanderbilt Initiative in Data-Intensive Astrophysics (VIDA). This work has made use of data from the European Space Agency (ESA) mission {\it Gaia} (\url{https://www.cosmos.esa.int/gaia}), processed by the {\it Gaia} Data Processing and Analysis Consortium (DPAC,
\url{https://www.cosmos.esa.int/web/gaia/dpac/consortium}). Funding for the DPAC has been provided by national institutions, in particular the institutions participating in the {\it Gaia} Multilateral Agreement. \\
This work was conducted at the University of California, San Diego, which was built on the unceded territory of the Kumeyaay Nation, whose people continue to maintain their political sovereignty and cultural traditions as vital members of the San Diego community.

\facilities{Gemini:South}
\software{emcee \citep{2013PASP..125..306F}, SMART \citep{2021ApJS..257...45H}, SME \& PySME \citep{1996A&AS..118..595V,2017A&A...597A..16P}, MOOG \citep{1973ApJ...184..839S}, SciPy \citep{2020SciPy-NMeth}, NumPy \citep{harris2020array}, matplotlib \citep{Hunter:2007}, astropy \citep{astropy:2013, astropy:2018, astropy:2022}, corner \citep{corner}, pysynphot \citep{2013ascl.soft03023S}}

\newpage
\appendix
\restartappendixnumbering
\section{Model grids and priors for spectral fitting}

\begin{deluxetable*}{cccccccccc}[h]

\tablecaption{Priors for determination of stellar and telluric parameters \label{tab:priors}}
\tabletypesize{\footnotesize}
\tablewidth{0pt}
\tablehead{\colhead{Target}    & \CellWithForceBreak{$T_\mathrm{eff}$ \\ (K)}    & \CellWithForceBreak{$\log{g}$ \\ (cgs)}      & \colhead{$\mathrm{[M/H]}$}     & \CellWithForceBreak{$v\sin{i}$\tablenotemark{*} \\ (km\,s$^{-1}$)}    & \CellWithForceBreak{$RV$\tablenotemark{*} \\ (km\,s$^{-1}$)} & \colhead{$\alpha$\tablenotemark{*}} & \colhead{$pwv$\tablenotemark{*}} & \colhead{LSF\tablenotemark{*}} & \CellWithForceBreak{Noise inflation \\ factor\tablenotemark{*} (N)} }
\startdata
HR 2562 & (6000, 7500) & (3.9, 5.1) & (-0.5, 0.5) & \multirow{5}{*}{(0,100)} & \multirow{5}{*}{(-100,100)} & \multirow{5}{*}{(0.0, 2.0)} & \multirow{5}{*}{(0.5, 5.0)} & \multirow{5}{*}{(0.1, 3.0)} & \multirow{5}{*}{(0.0, 5.0)} \\
AB Pic & (4000, 6000) & (3.5, 5.1) & (-0.5, 0.5) & & & & & &  \\
PZ Tel & (4500, 6500) & (3.5, 5.1) & (-0.5, 0.5) & & & & & &  \\
$\beta$ Pic & (7000, 9000) & (3.5, 4.9) & (-0.5, 0.5) & & & & & &  \\
YSES 1 & (4000, 5000) & (3.4, 5.1) & (-0.5, 0.5) & & & & & &  \\
\enddata
\tablenotetext{*}{All five targets use the same priors for these parameters}
\end{deluxetable*}

\begin{deluxetable}{ccccccccccc}
\tabletypesize{\scriptsize}
\tablecaption{Model grids used in this work \label{tab:grids}}
\tablewidth{0pt}
\tablehead{\colhead{Model grid}  & \colhead{$T_\mathrm{eff}$ (K)}  & \colhead{$\Delta T_\mathrm{eff}$ (K)}  & \colhead{$\log{g}$\tablenotemark{a}}  & \colhead{$\Delta\log{g}$\tablenotemark{a}} & \colhead{$\mathrm{[M/H]}$} & \colhead{$\Delta\mathrm{[M/H]}$} & \colhead{$\mathrm{[C/H]}$} & \colhead{$\Delta\mathrm{[C/H]}$} & \colhead{$\mathrm{[O/H]}$} & \colhead{$\Delta\mathrm{[O/H]}$}  } 

\startdata
\textit{PHOENIX}\tablenotemark{b} & (2300, 12000) & 500 & (0.0, 6.0) & 0.5 & (-4.0, 1.0) & 0.5 & - & - & - & - \\ \hline
\textit{MARCS}--\textit{C/O} & (6600, 6700) & 100 & (4.0, 5.0) & 0.5 & +0.10 & - & (-0.3, 0.5) & 0.1 & (-0.3, 0.7) & 0.1 \\
(HR 2562) & \\ \hline
\textit{MARCS}--\textit{C/O} &(4900, 5100) & 100 & (4.0, 4.5) & 0.5 & -0.07 & - & (-0.4, 0.5) & 0.1 & (-0.4, 0.7) & 0.1 \\
(AB Pic)  & \\ \hline
\textit{MARCS}--\textit{C/O}  & (4800, 5200) & 200 & (3.5,4.5) & 0.5 & -0.17 & - & (-0.4, 0.4) & 0.1 & (-0.4, 0.4) & 0.1 \\
\textit{MARCS}--\textit{C/O} \tablenotemark{c}  & (4800, 5200) & 200 & (3.5,4.5) & 0.5 & -0.17 & - & (-0.1, 0.1) & 0.1 & (-0.2, 0.8) & 0.1 \\
(PZ Tel) & \\ \hline
\textit{MARCS}--\textit{C/O} & (4600, 4800) & 100 & (4.0, 4.5) & 0.5 & -0.02 & - & (-0.4, 0.4) & 0.1 & (-0.5, 0.5) & 0.1 \\
(YSES 1) & \\ \hline
\textit{ATLAS}--\textit{C/O} & (7800, 8200) & 200 & (3.5, 4.5) & 0.5 & +0.01 & - & (-0.4, 0.4) & 0.1 & (-0.4, 1.0) & 0.1 \\
($\beta$ Pic)\tablenotemark{d} & \\ \hline
\textit{MARCS}--\textit{C/O} & (7000, 7400) & 200 & (4.0, 4.5) & 0.5 & -0.03 & - & (-0.4, 0.4) & 0.1 & (-0.4, 0.8) & 0.1 \\
(51 Eridani) \\
\enddata
\tablenotetext{a}{Surface gravity ($g$) in cgs units} 
\tablenotetext{b}{Standard grid created using stellar atmospheric code from \citet{10.1051/0004-6361/201219058}}
\tablenotetext{c}{Extended grid (limited in carbon) made to explore higher oxygen abundances}
\tablenotetext{d}{Custom abundances of specific elements, refer to text in section \ref{subsub:specfit}}
\end{deluxetable}

\newpage
\restartappendixnumbering
\section{Stellar Atmospheric Parameters Spectral Fit Plots}
\begin{figure}[h]
    \centering
    \includegraphics[width=1.0\linewidth]{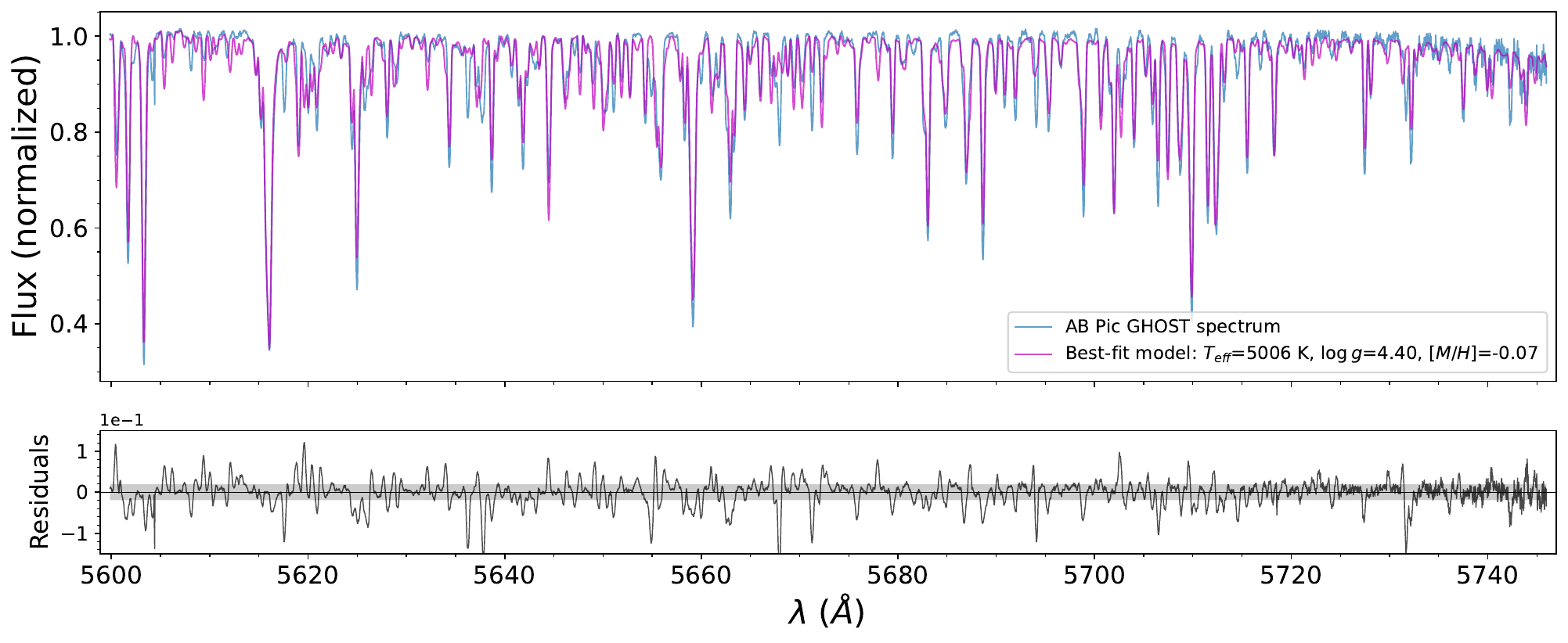}
    \caption{Best-fit PHOENIX model to the GHOST spectrum for the target AB Pic (cyan), shown for GHOST order 61. This model has $T_\mathrm{eff}$ = 5006 K, $\log{g}$ = 4.40, $\mathrm{[M/H]}$ = -0.07 (magenta). The residuals between the data and the model are plotted in black and other noise limits are shown in grey}
    \label{fig:atmosabpic}
\end{figure}
\begin{figure}
    \centering
    \includegraphics[width=1.0\linewidth]{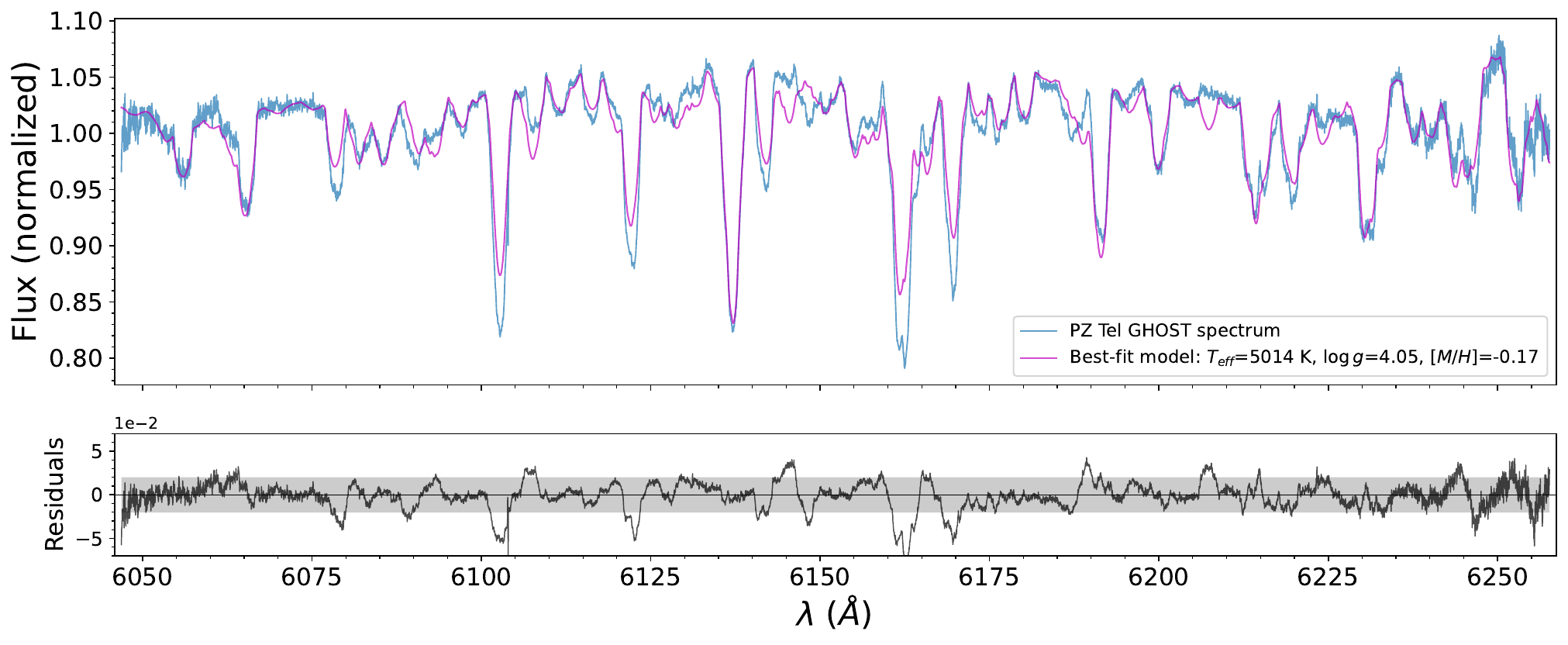}
    \caption{Best-fit PHOENIX model to the GHOST spectrum for the target PZ Tel (cyan), shown for GHOST order 56. This model has $T_\mathrm{eff}$ = 5014 K, $\log{g}$ = 4.05, $\mathrm{[M/H]}$ = -0.17 (magenta). The residuals between the data and the model are plotted in black and other noise limits are shown in grey}
    \label{fig:atmospztel}
\end{figure}

\begin{figure}
    \centering
    \includegraphics[width=1.0\linewidth]{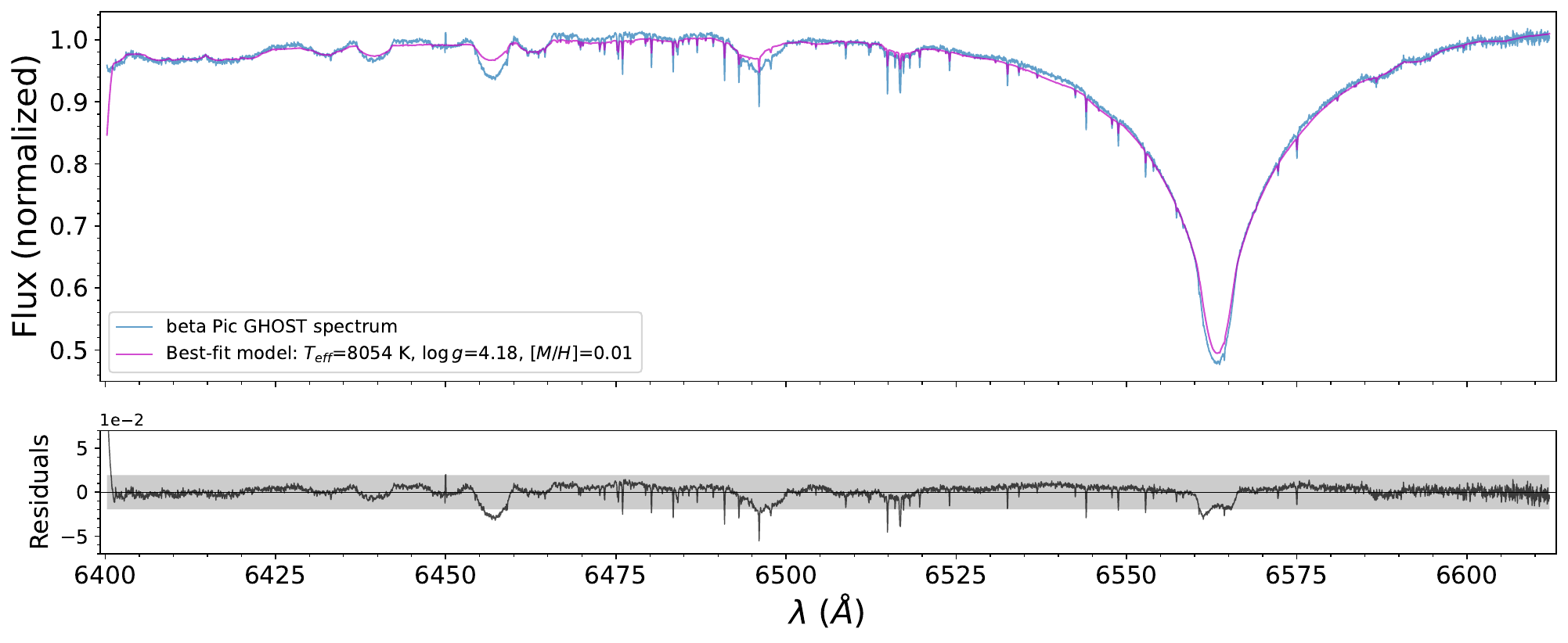}
    \caption{Best-fit PHOENIX model to the GHOST spectrum for the target $\beta$ Pic (cyan), shown for GHOST order 53. This model has $T_\mathrm{eff}$ = 8054 K, $\log{g}$ = 4.18, $\mathrm{[M/H]}$ = 0.01 (magenta). The residuals between the data and the model are plotted in black and other noise limits are shown in grey}
    \label{fig:atmosbetapic}
\end{figure}
\begin{figure}
    \centering
    \includegraphics[width=1.0\linewidth]{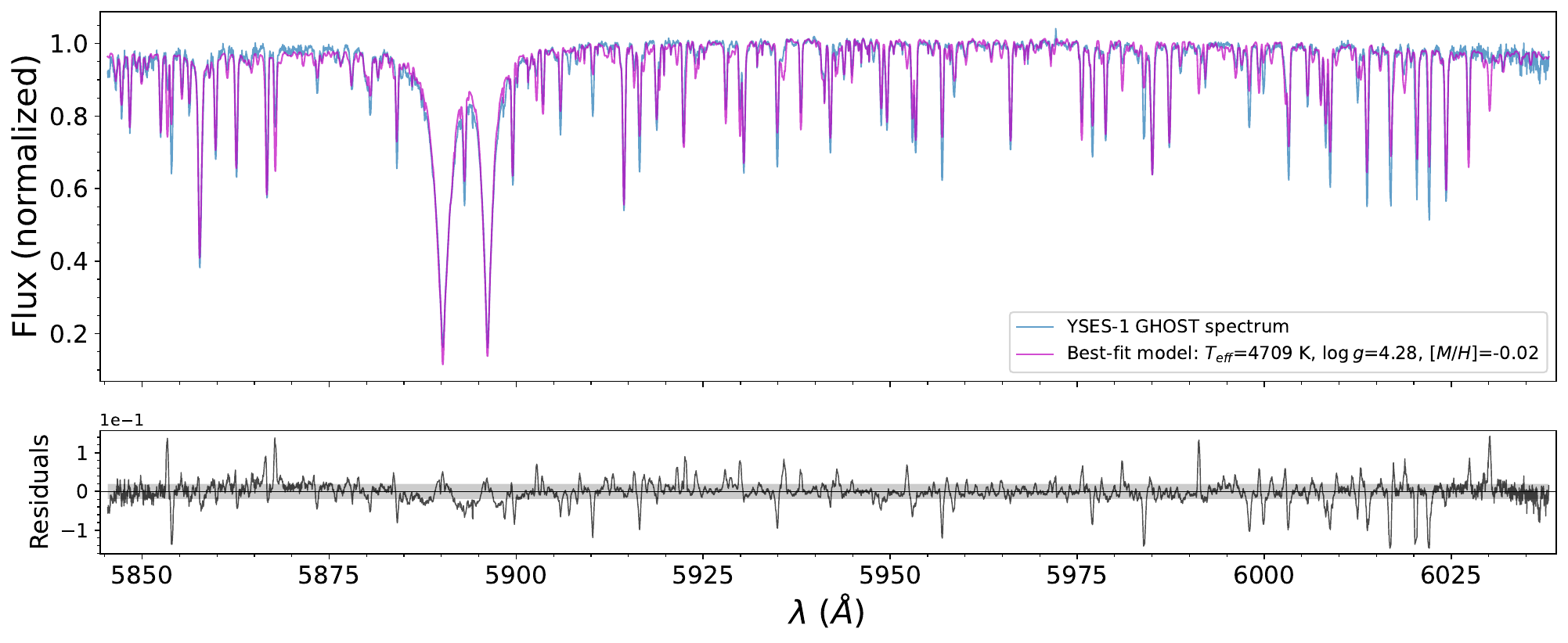}
    \caption{Best-fit PHOENIX model to the GHOST spectrum for the target YSES 1 (cyan), shown for GHOST order 58. This model has $T_\mathrm{eff}$ = 4709 K, $\log{g}$ = 4.28, $\mathrm{[M/H]}$ = -0.02 (magenta). The residuals between the data and the model are plotted in black and other noise limits are shown in grey}
    \label{fig:atmosyses1}
\end{figure}

\begin{figure}
    \centering
    \includegraphics[width=0.45\linewidth]{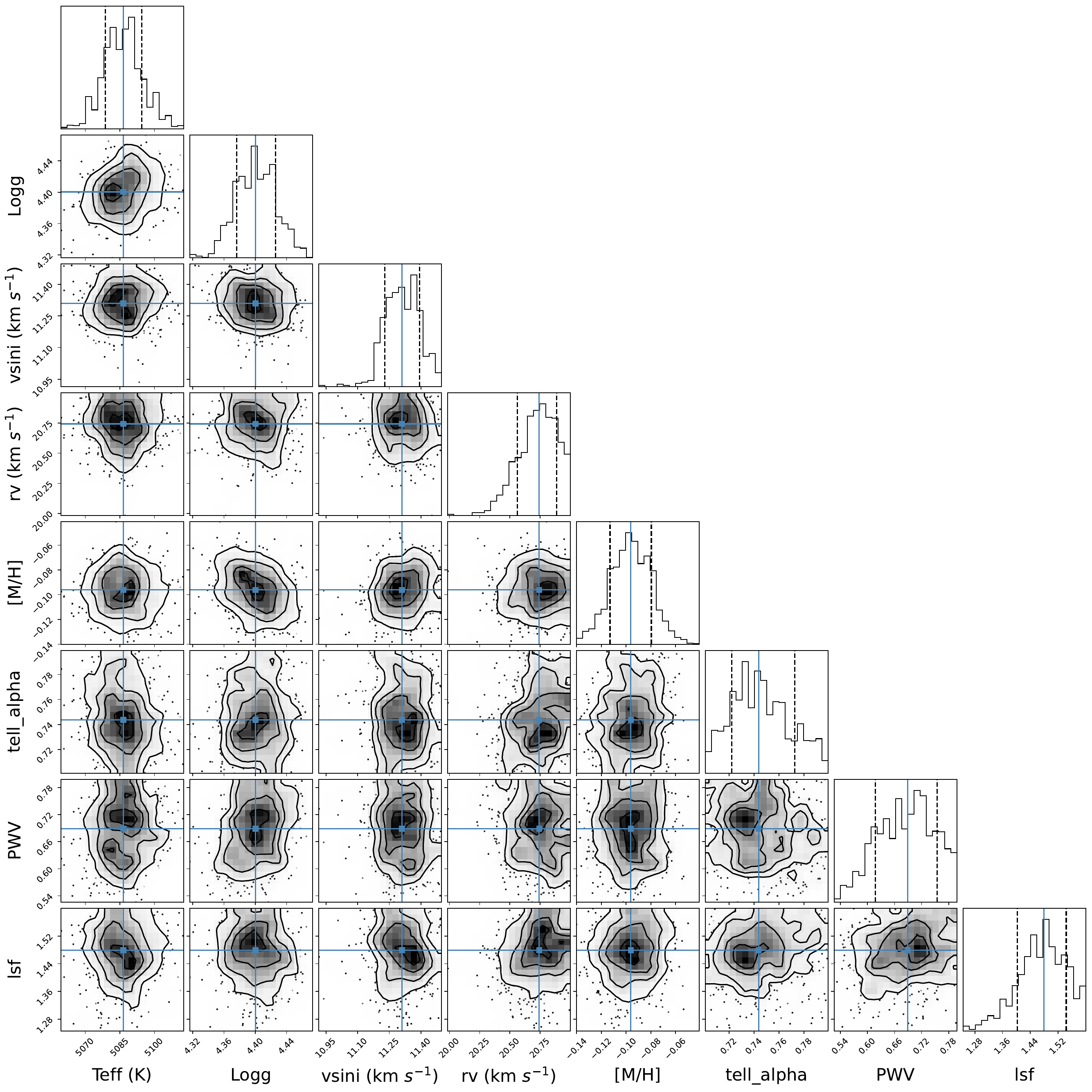}
    \caption{Corner plot for one of the MCMC runs for a multi-order fit of the PHOENIX grid to the spectrum of AB Pic. The marginalized posteriors are shown along the diagonal. The blue lines represent the 50 percentile, and the dotted lines represent the 16 and 84 percentiles. The subsequent covariances between all the parameters are in the corresponding 2-D histograms. This MCMC run gives best-fit $T_\mathrm{eff}$ = 5086 $\pm$ 8 K, $\log{g}$ = 4.40 $\pm$ 0.02, [M/H] = -0.10 $\pm$ 0.02 dex.}
    \label{fig:coatmoabpic}
\end{figure}

\begin{figure}
    \centering
    \includegraphics[width=0.45\linewidth]{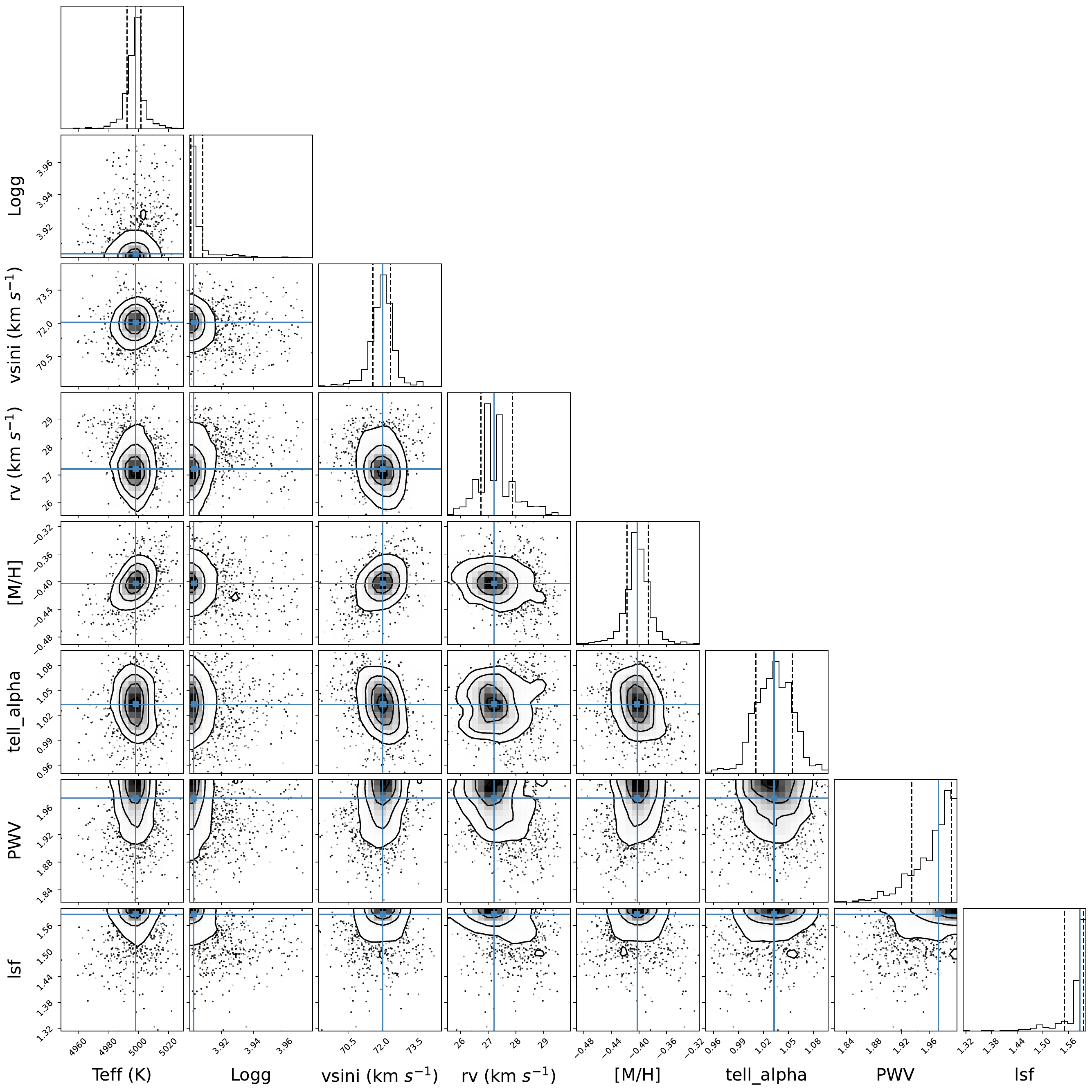}
    \caption{Corner plot for one of the MCMC runs for a multi-order fit of the PHOENIX grid to the spectrum of PZ Tel. The marginalized posteriors are shown along the diagonal. The blue lines represent the 50 percentile, and the dotted lines represent the 16 and 84 percentiles. The subsequent covariances between all the parameters are in the corresponding 2-D histograms. This fit was among those used to determine the best-fit temperature after constraining $\log{g}$. The metallicity is poorly constrained by the fitted orders. This run has a best-fit $T_\mathrm{eff}$ = 4998 $\pm$ 6 K.}
    \label{fig:coatmopztel}
\end{figure}

\begin{figure}
    \centering
    \includegraphics[width=0.45\linewidth]{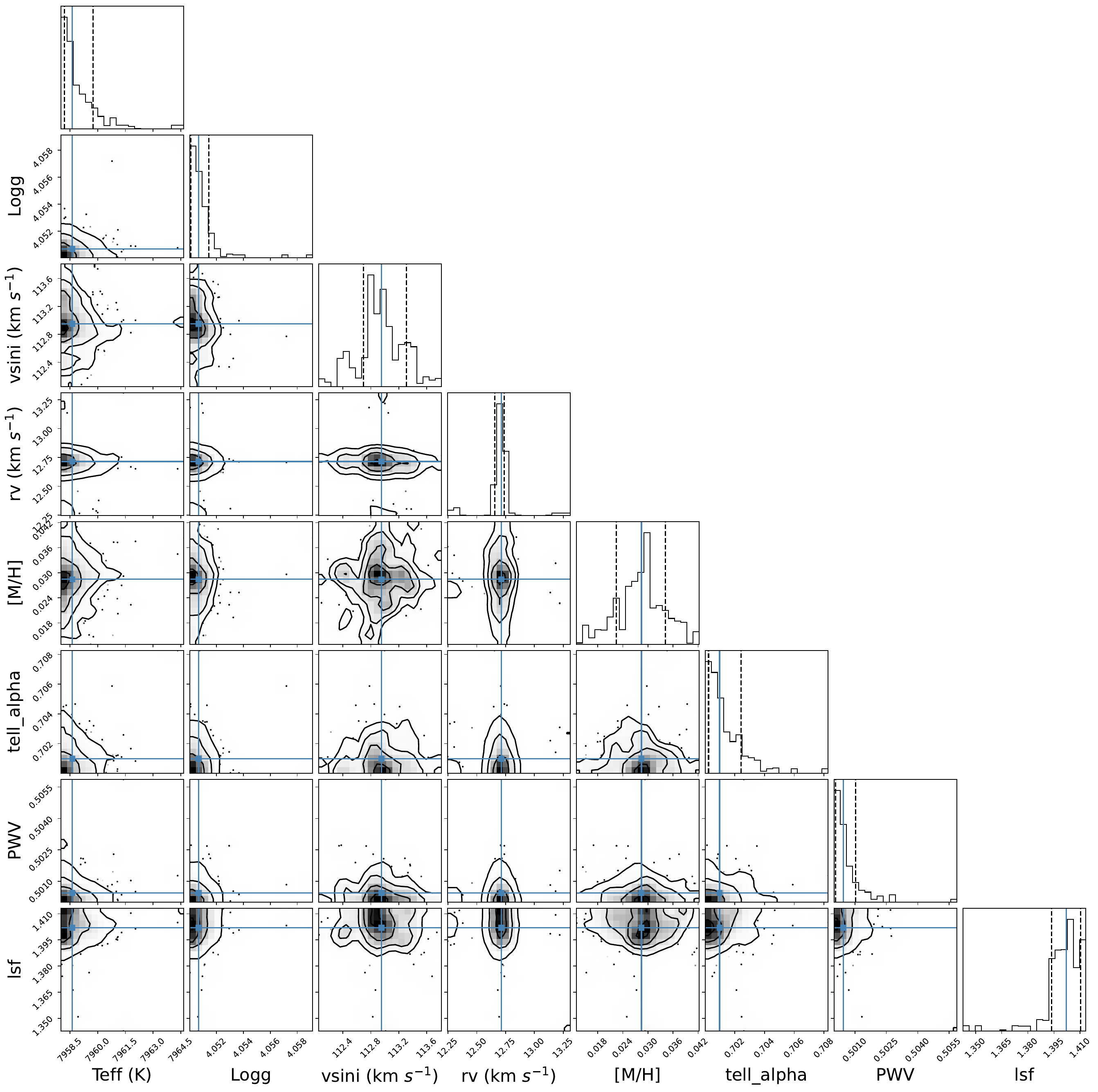}
    \caption{Corner plot for one of the MCMC runs for a multi-order fit of the PHOENIX grid to the spectrum of $\beta$ Pic. The marginalized posteriors are shown along the diagonal. The blue lines represent the 50 percentile, and the dotted lines represent the 16 and 84 percentiles. The subsequent covariances between all the parameters are in the corresponding 2-D histograms. This fit was among those used to determine the best-fit metallicity after constraining $T_\mathrm{eff}$ and $\log{g}$. This run has a best-fit [M/H] = 0.03 $\pm$ 0.01 dex.}
    \label{fig:coatmobetapic}
\end{figure}

\begin{figure}
    \centering
    \includegraphics[width=0.45\linewidth]{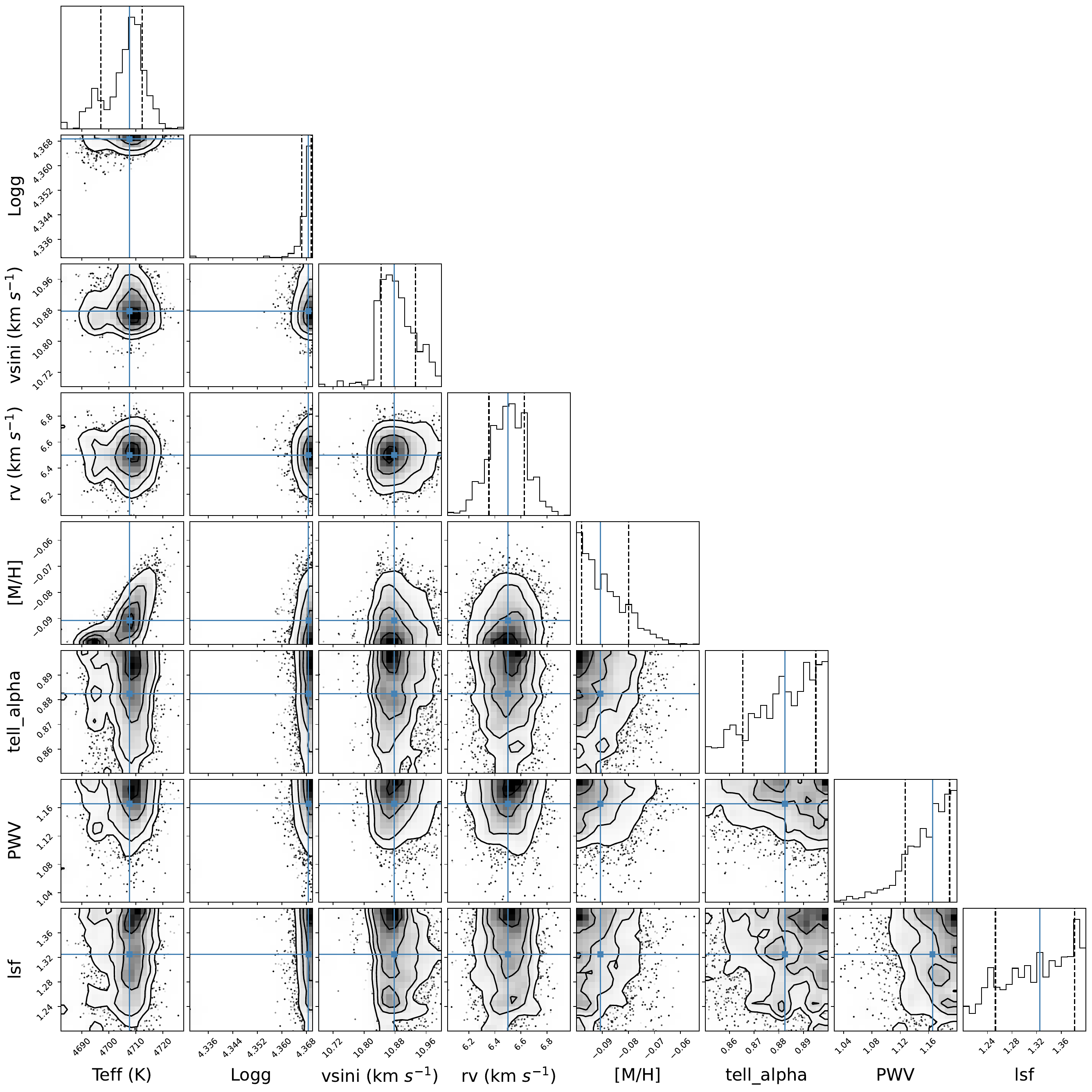}
    \caption{Corner plot for a single-order fit to order 58 (with NaI D doublet) of the PHOENIX grid to the spectrum of YSES 1. The marginalized posteriors are shown along the diagonal. The blue lines represent the 50 percentile, and the dotted lines represent the 16 and 84 percentiles. The subsequent covariances between all the parameters are in the corresponding 2-D histograms. This fit was among those used to determine the best-fit temperature after constraining $\log{g}$, and a weak constraint on [M/H]. This run has a best-fit temperature = 4708 $\pm$ 11 K.}
    \label{fig:coatmoyses1}
\end{figure}

\restartappendixnumbering
\section{Carbon and Oxygen Abundances Spectral Fit Plots}

\begin{figure}[H]
    \centering
    \plottwo{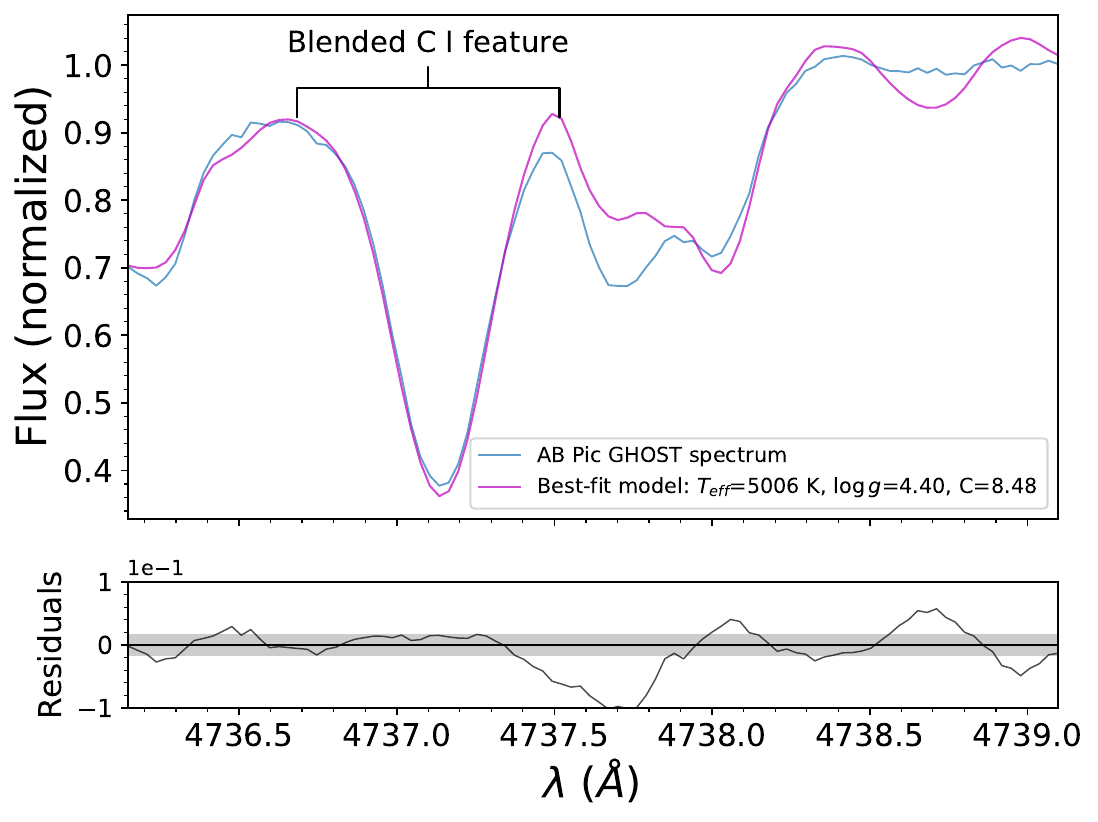}{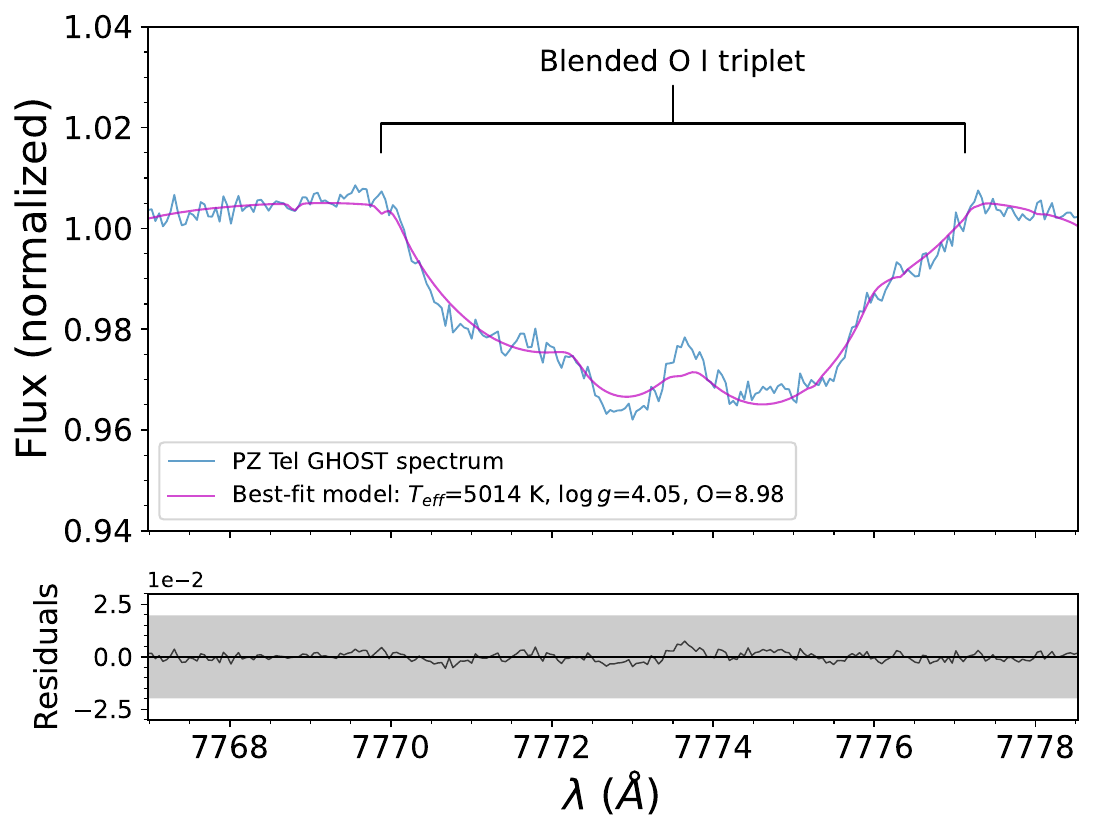}
    \caption{(\textit{Left)} Best-fit \textit{MARCS-C/O} model (magenta) to the GHOST spectrum of the star AB Pic (cyan), shown for the blended C I feature at 4737 \r{A}. The best-fit $\log{\epsilon_C}$ = 8.48. \\
    (\textit{Right}) Best-fit \textit{MARCS-C/O} model (magenta) to the GHOST spectrum of the star PZ Tel (cyan), shown for the blended O I triplet feature at 7771--75 \r{A}. The best-fit $\log{\epsilon_O}$ = 8.98, which after applying NLTE corrections, comes out to $\log{\epsilon_O}$ = 8.95. In both plots, the residuals between the data and the model are plotted in black and other noise limits are shown in grey.}
    \label{fig:abpicpztelspecco}
\end{figure}

\begin{figure}[H]
    \centering
    \plottwo{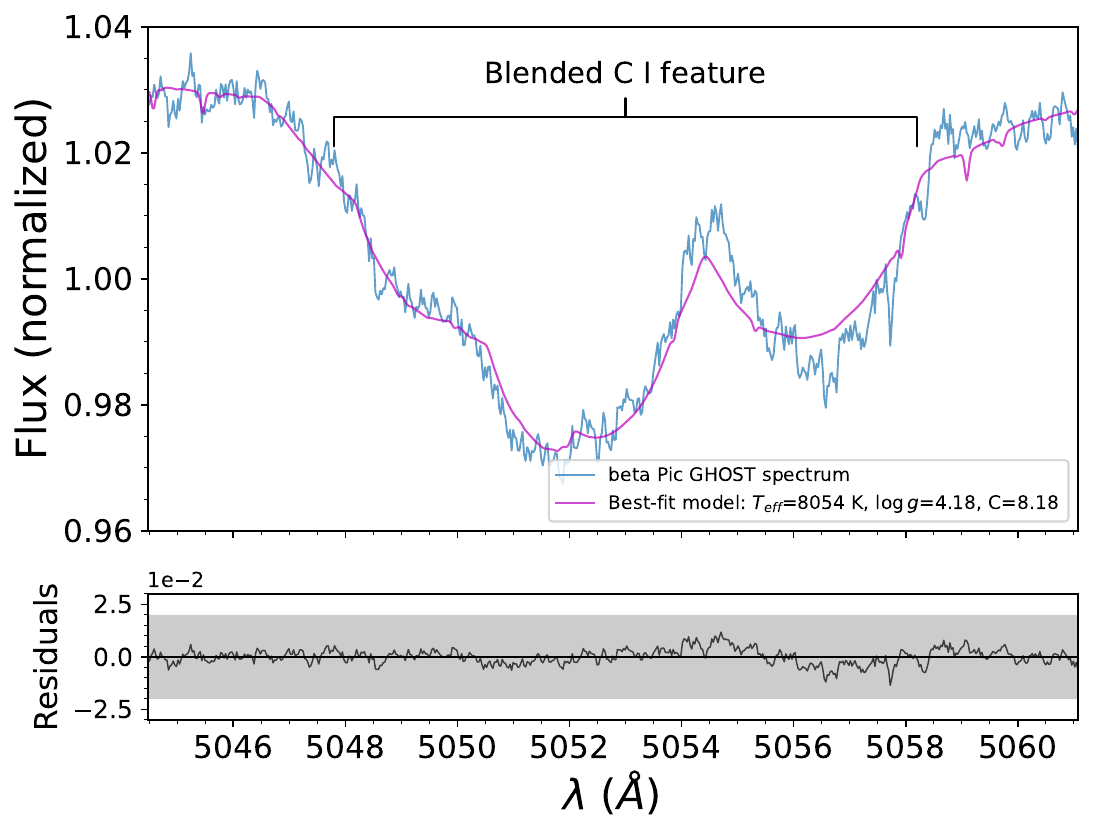}{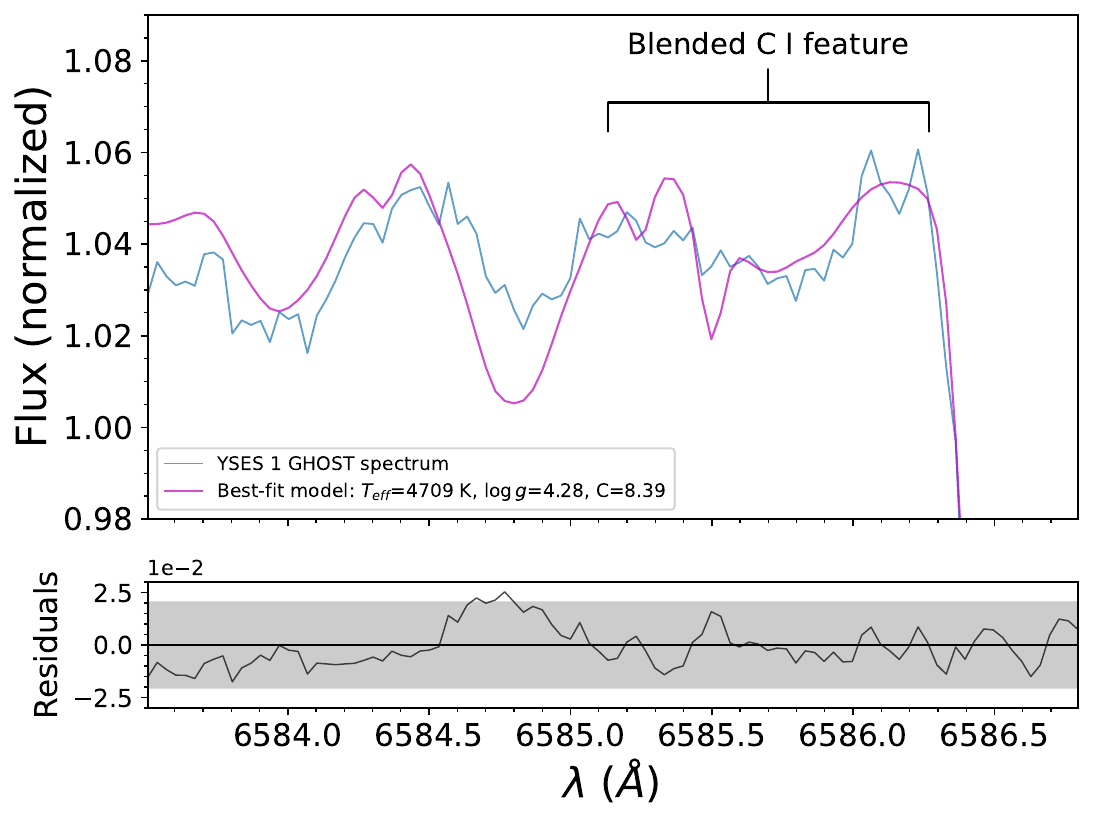}
    \caption{(\textit{Left)} Best-fit \textit{ATLAS-C/O} model (magenta) to the GHOST spectrum of the star $\beta$ Pic (cyan), shown for the blended C I feature at 5052 \r{A}. The best-fit $\log{\epsilon_C}$ = 8.18. \\
    (\textit{Right}) Best-fit \textit{MARCS-C/O} model (magenta) to the GHOST spectrum of the star YSES 1 (cyan), shown for the blended C I feature at 6585--6586 \r{A}. The best-fit $\log{\epsilon_C}$ = 8.39. In both plots, the residuals between the data and the model are plotted in black and other noise limits are shown in grey.}
    \label{fig:betapicyses1specco}
\end{figure}

\begin{figure}[H]
    \centering
    \includegraphics[width=0.45\linewidth]{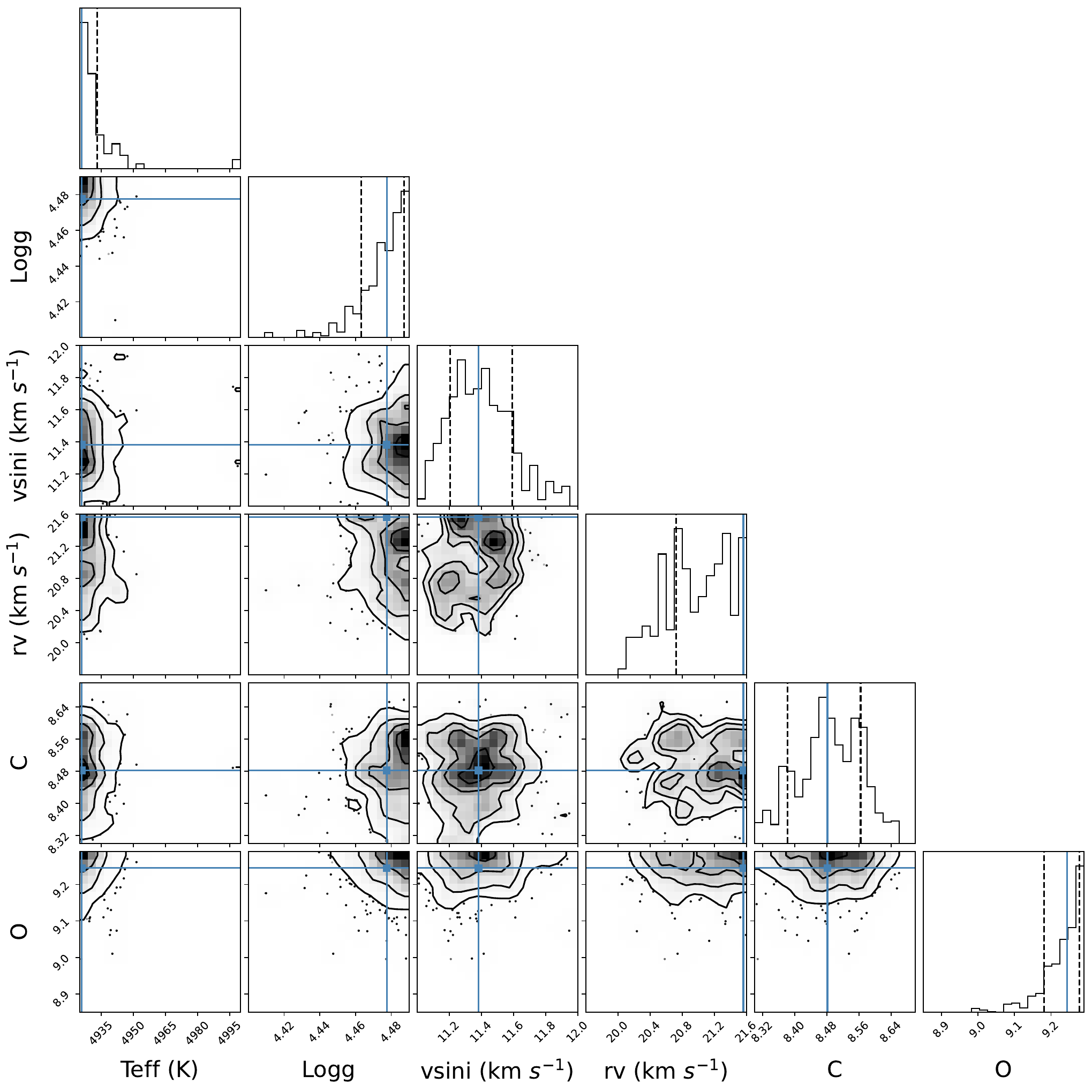}
    \caption{Corner plot for \textit{MARCS}--\textit{C/O} grid fit to spectral orders with carbon features for AB Pic. The marginalized posteriors are shown along the diagonal. The blue lines represent the 50 percentile, and the dotted lines represent the 16 and 84 percentiles. The subsequent covariances between all the parameters are in the corresponding 2-D histograms. We obtain a best-fit $\log{\epsilon_C}$ = 8.48 $\pm$ 0.10. The oxygen is not well-constrained while fitting the carbon orders.}
    \label{fig:coabpic}
\end{figure}

\begin{figure}[H]
    \centering
    \includegraphics[width=0.45\linewidth]{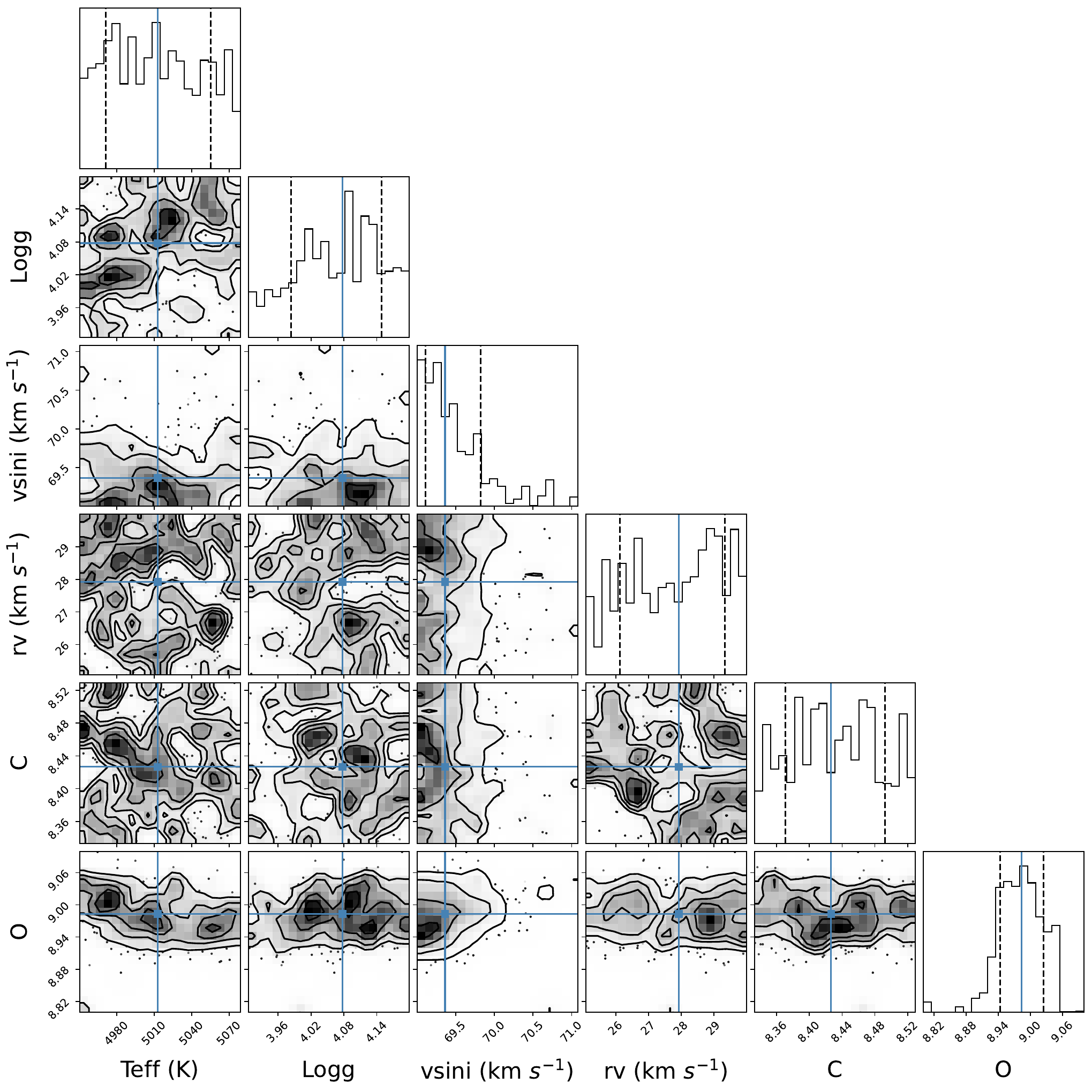}
    \caption{Corner plot for \textit{MARCS}--\textit{C/O} grid fit to spectral order with the O I triplet at 7771--75 \r{A}. The marginalized posteriors are shown along the diagonal. The blue lines represent the 50 percentile, and the dotted lines represent the 16 and 84 percentiles. The subsequent covariances between all the parameters are in the corresponding 2-D histograms. We obtain a best-fit $\log{\epsilon_O}$ = 8.98 $\pm$ 0.04, corresponding to $\log{\epsilon_O}$ = 8.95 $\pm$ 0.04 after NLTE correction. The carbon is not well-constrained fitting this order.}
    \label{fig:copztel}
\end{figure}

\begin{figure}[H]
    \centering
    \includegraphics[width=0.45\linewidth]{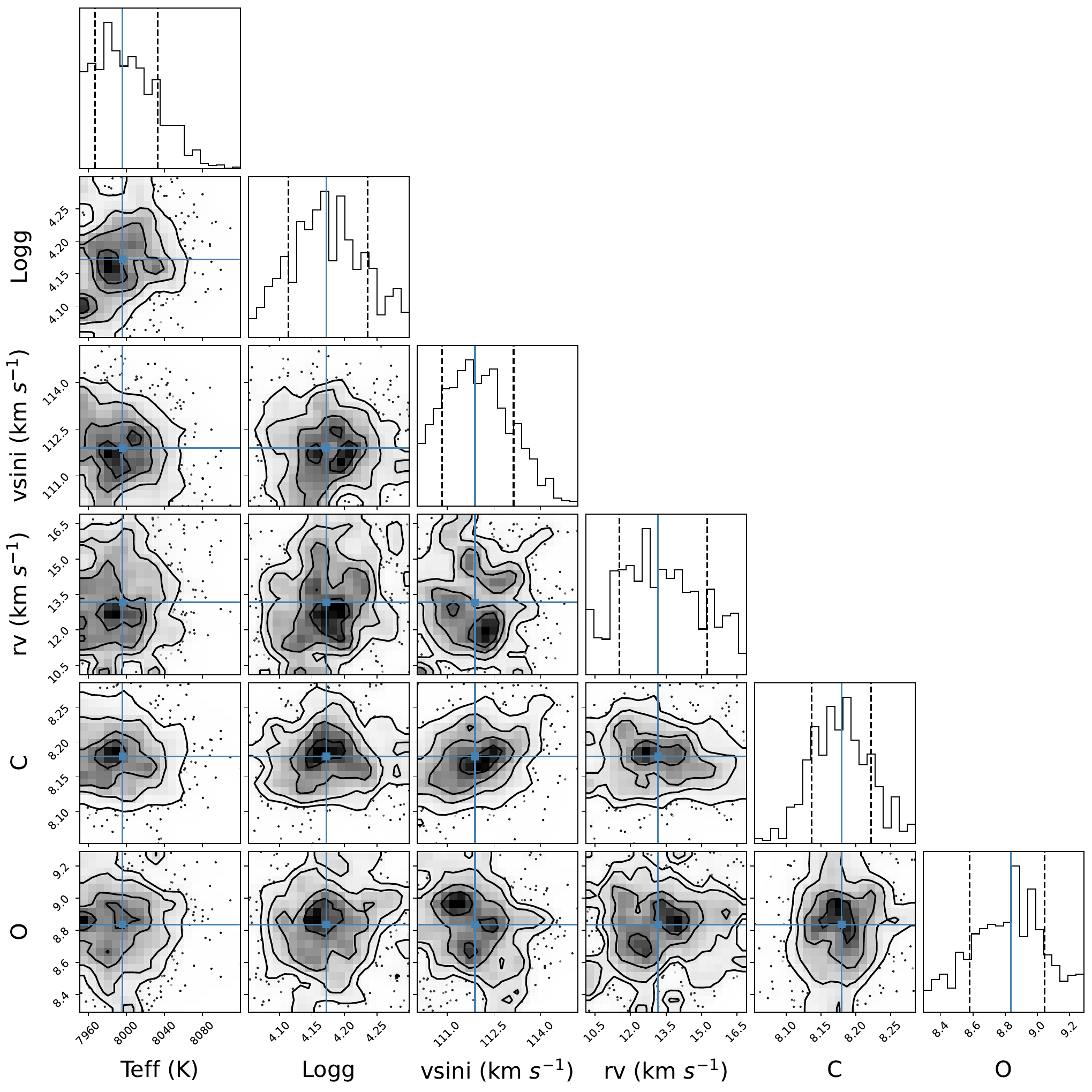}
    \caption{Corner plot for \textit{ATLAS}--\textit{C/O} grid fit to spectral orders with carbon features for $\beta$ Pic. The marginalized posteriors are shown along the diagonal. The blue lines represent the 50 percentile, and the dotted lines represent the 16 and 84 percentiles. The subsequent covariances between all the parameters are in the corresponding 2-D histograms. We obtain a best-fit $\log{\epsilon_C}$ = 8.18 $\pm$ 0.06. The oxygen is not well-constrained while fitting the carbon orders.}
    \label{fig:cobetapic}
\end{figure}

\begin{figure}[H]
    \centering
    \includegraphics[width=0.45\linewidth]{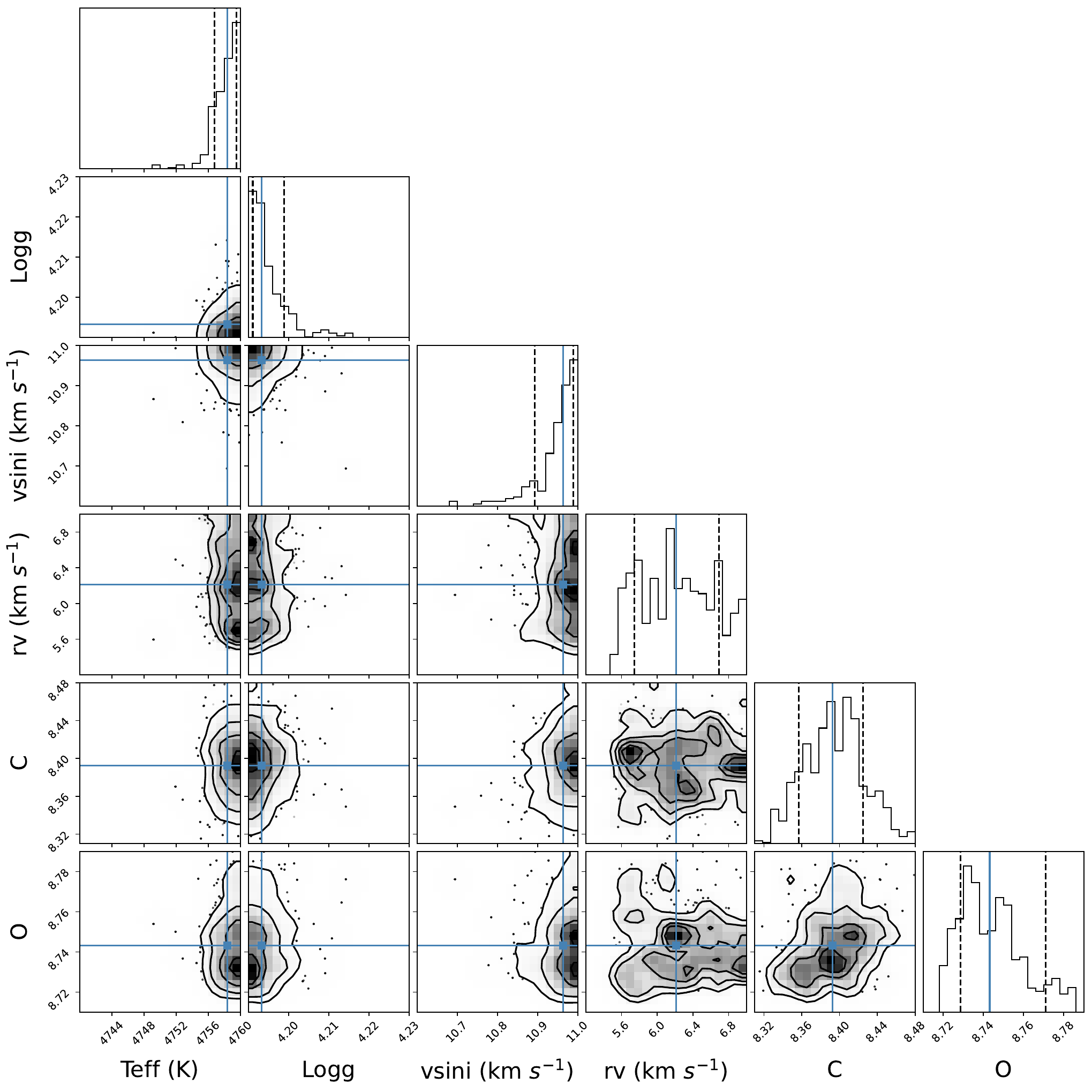}
    \caption{Corner plot for \textit{MARCS}--\textit{C/O} grid fit to spectral orders with carbon features for YSES 1. The marginalized posteriors are shown along the diagonal. The blue lines represent the 50 percentile, and the dotted lines represent the 16 and 84 percentiles. The subsequent covariances between all the parameters are in the corresponding 2-D histograms. We obtain a best-fit $\log{\epsilon_C}$ = 8.39 $\pm$ 0.04. The oxygen is not well-constrained while fitting the carbon orders.}
    \label{fig:coyses1}
\end{figure}

\bibliography{main}

\end{document}